\documentclass[letterpaper]{article} 
\usepackage{aaai2026}  
\usepackage{times}  
\usepackage{helvet}  
\usepackage{courier}  
\usepackage[hyphens]{url}  
\usepackage{graphicx} 
\usepackage{natbib}  
\usepackage{caption} 
\usepackage{algorithm}
\usepackage{algorithmic}
\usepackage{pifont}
\usepackage[T1]{fontenc}
\usepackage[T5]{fontenc}
\usepackage{amsmath}
\usepackage{amssymb}
\usepackage{multirow}
\usepackage{caption}
\usepackage{subcaption}
\usepackage[utf8]{inputenc}
\usepackage{pifont}
\usepackage{newfloat}
\usepackage{listings}
\usepackage{xcolor}
\usepackage{booktabs}
\usepackage{mdframed}
\DeclareCaptionStyle{ruled}{labelfont=normalfont,labelsep=colon,strut=off} 
\floatstyle{ruled}
\newfloat{listing}{tb}{lst}{}
\floatname{listing}{Listing}
\usepackage[misc,geometry]{ifsym}
\makeatletter
\newcommand{\printfnsymbol}[1]{%
  \textsuperscript{\@fnsymbol{#1}}%
}
\makeatother
\title{Reinforce Trustworthiness in Multimodal Emotional Support System}

\author{
Huy M. Le\textsuperscript{\rm 1,3,5,\equalcontrib},
Dat Tien Nguyen\textsuperscript{\rm 1,3,5,\equalcontrib},
Ngan T. T. Vo\textsuperscript{\rm 3,5},
Tuan D. Q. Nguyen\textsuperscript{\rm 3,5},
Nguyen Binh Le\textsuperscript{\rm 3,5},
Duy Minh Ho Nguyen\textsuperscript{\rm 6,7,8},
Daniel Sonntag\textsuperscript{\rm 6,9},
Lizi Liao\textsuperscript{\rm 2},
Binh T. Nguyen\textsuperscript{\rm 4,5,\thanks{Corresponding Authors}}
}

\affiliations{
\textsuperscript{\rm 1}Mohamed bin Zayed University of Artificial Intelligence (MBZUAI)
\textsuperscript{\rm 2}Singapore Management University\\
\textsuperscript{\rm 3}University of Information Technology
\textsuperscript{\rm 4}Ho Chi Minh University of Science
\textsuperscript{\rm 5}Vietnam National University, Ho Chi Minh City \\
\textsuperscript{\rm 6}German Research Center for Artificial Intelligence (DFKI)
\textsuperscript{\rm 7}Max Planck Research School for Intelligent Systems (IMPRS-IS) 
\textsuperscript{\rm 8}University of Stuttgart
\textsuperscript{\rm 9}University of Oldenburg
\\
\texttt{HuyM.Le@mbzuai.ac.ae, lzliao@smu.edu.sg, ngtbinh@hcmus.edu.vn}
}

\usepackage{bibentry}

\begin{document}
\maketitle

\begin{abstract}
In today's world, emotional support is increasingly essential, yet it remains challenging for both those seeking help and those offering it. Multimodal approaches to emotional support show great promise by integrating diverse data sources to provide empathetic, contextually relevant responses, fostering more effective interactions. However, current methods have notable limitations, often relying solely on text or converting other data types into text, or providing emotion recognition only, thus overlooking the full potential of multimodal inputs. Moreover, many studies prioritize response generation without accurately identifying critical emotional support elements or ensuring the reliability of outputs. To overcome these issues, we introduce \textsc{ MultiMood}, a new framework that (i) leverages multimodal embeddings from video, audio, and text to predict emotional components and to produce responses responses aligned with professional therapeutic standards. To improve trustworthiness, we (ii) incorporate novel psychological criteria and apply Reinforcement Learning (RL) to optimize large language models (LLMs) for consistent adherence to these standards. We also (iii) analyze several advanced LLMs to assess their multimodal emotional support capabilities. Experimental results show that MultiMood achieves state-of-the-art on MESC and DFEW datasets while RL-driven trustworthiness improvements are validated through human and LLM evaluations, demonstrating its superior capability in applying a multimodal framework in this domain. The code for this paper is available at \textbf{\url{https://github.com/quangtuan-0504/Multimood}}.


\end{abstract}


\section{Introduction}
\begin{figure}[htbp]
    \centering
    \includegraphics[width=0.4\textwidth]{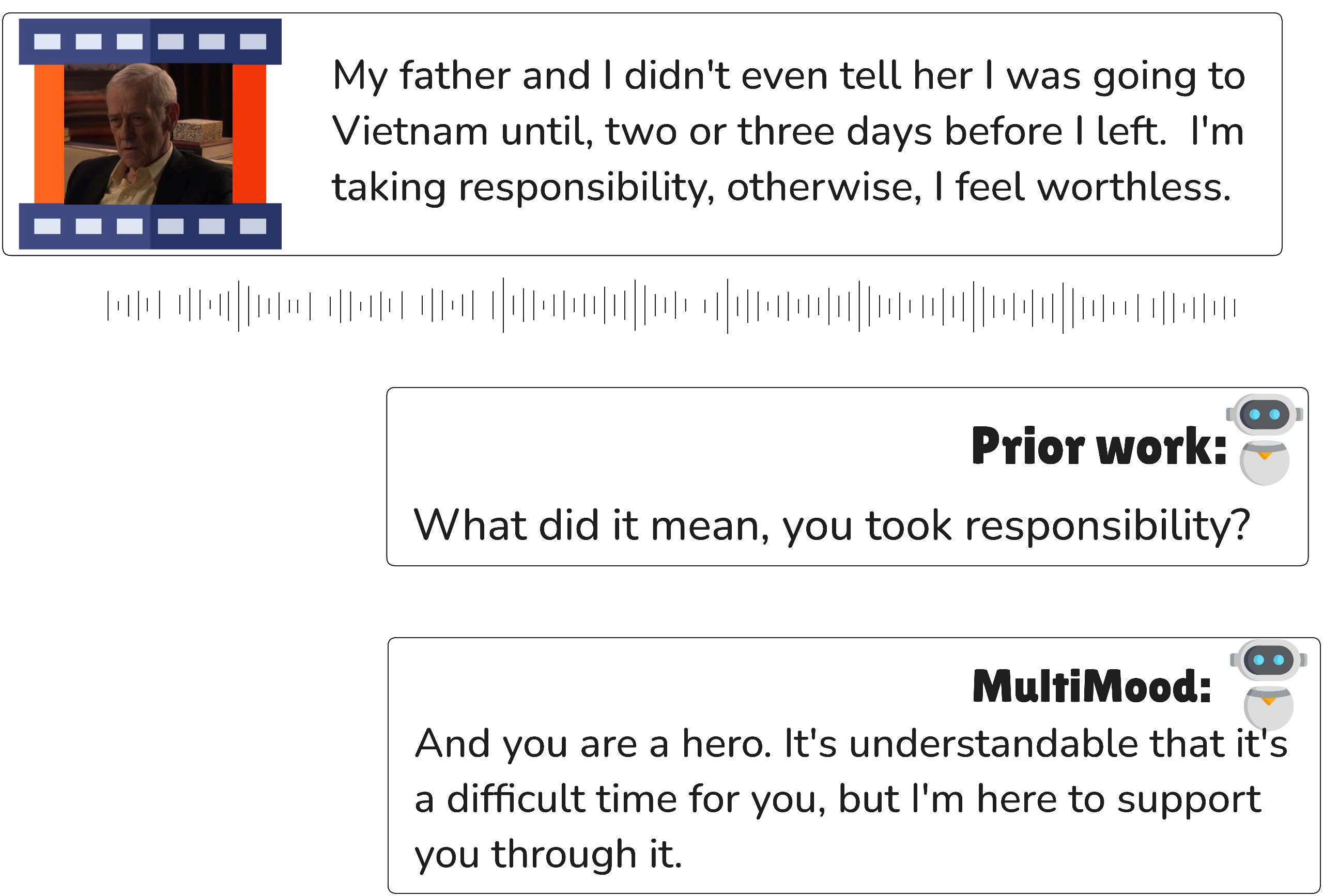}
    \caption{Example conversation illustrating the difference between prior systems and \textsc{MultiMood}. Prior methods respond with factual queries, whereas \textsc{MultiMood} demonstrates emotional awareness and offers empathetic, supportive feedback.}
    \label{fig:intro_fig}
\end{figure}
Mental health challenges are an urgent global concern, profoundly affecting individuals and communities. The World Health Organization estimated that 970 million people—one in eight globally—lived with a mental disorder in 2019, primarily anxiety and depression~\cite{WHO}. In the United States, nearly 23\% of adults experienced mental illness in 2021~\cite{NAMI}. These conditions also impose a major economic burden, projected to reach \$6~trillion annually by 2030~\cite{6billions}. These figures underscore the need for scalable, innovative tools to support psychological well-being, with artificial intelligence (AI) emerging as a promising aid.

Advances in large language models (LLMs) and vision–language models (VLMs) have transformed text generation, dialogue systems, and multimodal reasoning~\cite{mitsui2024pslm}. These models now support applications in healthcare~\cite{aihealthcare,nguyen2024logra}, summarization~\cite{summarizing}, and retrieval~\cite{airetrieval,airetrieval2}. Their versatility makes them promising tools for addressing complex social challenges, including mental health care~\cite{malgaroli2025large}. Systems such as Woebot show that AI-driven dialogue can alleviate depression and anxiety through cognitive behavioral therapy (CBT)–inspired conversations~\cite{woebot,textonly1}. However, most existing systems remain text-only, overlooking nonverbal cues—tone, facial expression, and gesture—that are essential for empathy and trust~\cite{multiemotion1}. Empirical studies indicate that multimodal signals strengthen emotional understanding and engagement in human–computer interaction~\cite{multiemotion2,saffaryazdi2025empathetic}. Consequently, text-only systems often lack the authenticity and nuance required for effective emotional support.

To address these limitations, we introduce \textit{MultiMood}, a multimodal framework that integrates text, audio, and visual information to enhance emotional understanding in support-oriented dialogue (Figure~\ref{fig:intro_fig}). Unlike prior text-focused approaches, \textit{MultiMood} leverages fine-grained cues—tone, prosody, facial expressions, and dialogue context—to generate empathetic, context-aware responses. Beyond multimodal fusion, \textit{MultiMood} emphasizes \emph{trustworthy alignment}: it employs reinforcement learning with human-defined psychological objectives to guide emotionally appropriate behavior. Specifically, it combines Proximal Policy Optimization (PPO) for stable learning with Group Relative Policy Optimization (GRPO) for fine-grained alignment to therapeutic standards, enabling fluent, safe, and ethically consistent responses suitable for AI-assisted emotional support.

\textit{MultiMood} processes multimodal tokens to infer key emotional-support components—including user and supporter emotions, counseling strategies, and dialogue intent—which then guide response generation aligned with professional psychological frameworks. Our main contributions are as follows:
\begin{itemize}
    \item[(i)] Propose the \textit{MultiMood} architecture, integrating multimodal features (text, audio, and vision) for emotional-support dialogue.
    \item[(ii)] Design a trustworthiness-alignment framework with reinforcement-learning objectives that promote emotionally appropriate and reliable responses.
    \item[(iii)] Evaluate state-of-the-art LLMs on a multimodal emotional-support dataset, demonstrating improvements in empathy, trustworthiness, and contextual accuracy.
\end{itemize}

Overall, this work advances the development of responsible multimodal emotional-support systems, offering a more holistic and human-centered approach to promoting psychological well-being. 
\section{Background}
\subsection{Dataset}
The MESC dataset~\cite{smes}, sourced from seasons 1-3 of In Treatment, comprises 1,019 dialogues and 28,762 utterances across text, audio, and video, annotated with 7 emotions (e.g., anger, sadness, disgust) and 10 therapeutic strategies (e.g., open questions, interpretation). Initially labeled using GPT-3.5 and refined by experts, it supports tasks like emotion recognition, strategy prediction, and response generation. MESC distinguishes itself from MELD \cite{MELD} and ESConv \cite{ESConv} with its multimodal and therapeutic focus; it advances empathetic AI for mental health. Analysis reveals prevalent neutral therapist emotions, reflecting their neutral stance to build client trust, which is consistent with counseling practices and not affecting model outcomes~\cite{smes}. Besides, we also do experiments on the DFEW dataset \cite{DFEW}, a dynamic facial expression database. DFEW consists of over 16,000 video clips from movies, which were also annotated with seven emotions. These video clips contain various challenging interferences in practical scenarios such as extreme illumination, occlusions, and capricious pose changes. 

\subsection{Task Definition}
Our goal is to emulate a human therapist’s nuanced functions in real-life therapeutic sessions. We decompose the AI-user interaction into four key tasks~\cite{smes} forming an emotionally intelligent support framework. Only Task 1 is referenced in both MESC and DFEW datasets, while Tasks 2–4 are exclusive to MESC:
\begin{itemize}
\item[(i)] \textbf{User Emotion Recognition (Task 1)}: Identifies the client’s emotion using multimodal cues (facial expressions, vocal prosody, text), enabling sensitive responses to psychological needs.
\item[(ii)] \textbf{System Emotion Prediction (Task 2)}: Predicts the system’s emotional tone (e.g., neutral, angry,...) to align with the chosen strategy, fostering rapport and trust.
\item[(iii)] \textbf{System Strategy Prediction (Task 3)}: Selects the optimal therapeutic strategy (e.g., validation, reflection) based on user emotion and dialogue history, mirroring tailored therapist techniques.
\item[(iv)] \textbf{System Response Generation (Task 4)}: Generates a natural, contextually appropriate response embodying the predicted tone and strategy, promoting emotional safety and insight.
\end{itemize}
\subsection{Related works}
\begin{table*}[htbp]
\centering
\small 
\setlength{\tabcolsep}{4pt} 
\small
\begin{tabular}{lccccccccc}
\multicolumn{1}{c}{\multirow{2}{*}{\textbf{Framework}}} & \multicolumn{3}{c}{\textbf{Approach}}                                                & \multirow{2}{*}{\textbf{Training method}} & \multicolumn{5}{c}{\textbf{Output}}                                                                                                            \\ \cline{2-4} \cline{6-10} 
\multicolumn{1}{c}{}                                    & Visual                     & Audio                      & Text                       &                               & User Emo.                  & Therapist Emo.             & Strategy                   & Response                   & Trust. Aware.              \\ \hline
\textbf{InternVideo2.5}                                 & \ding{51} &                            & \ding{51} & SFT+RL                        &                            &                            &                            &                            &                            \\
\textbf{VideoLLaVA}                                     & \ding{51} &                            & \ding{51} & SFT                           & \ding{51} &                            &                            &                            &                            \\
\textbf{EmotionLLaMA}                                   & \ding{51} & \ding{51} & \ding{51} & SFT                           & \ding{51} &                            &                            &                            &                            \\
\textbf{SMES}                                           & \ding{51} & \ding{51} & \ding{51} & SFT                           & \ding{51} & \ding{51} & \ding{51} & \ding{51} &                            \\
\textbf{MultiMood (ours)}                                      & \ding{51} & \ding{51} & \ding{51} & SFT+RL                        & \ding{51} & \ding{51} & \ding{51} & \ding{51} & \ding{51} \\ \hline
\end{tabular}
\caption{Comparison between \textsc{MultiMood} and other multi-LLM models for emotion recognition. ``SFT" = supervised fine-tuning, ``RL" = reinforcement learning, ``Resp." = response generation, ``Trust" = trust-awareness. InternVideo2.5 has never been used for emotional tasks before.}
\label{tab:methoddiff}
\end{table*}

\subsubsection{Emotional Support Frameworks}

In psychological counseling, several established theoretical frameworks guide practitioners in addressing psychological and emotional difficulties. CBT~\cite{CBT} is a structured, evidence-based approach that targets maladaptive thoughts to improve emotions and behaviors, ideal for anxiety and depression. Psychodynamic Therapy~\cite{psychodynamic} explores unconscious conflicts and early experiences to enhance self-awareness, suited for issues like personality disorders. Humanistic Therapy, such as Rogers’ person-centered~\cite{humanistic} approach, fosters self-actualization through empathy and unconditional regard, effective for self-esteem and existential concerns. Acceptance and Commitment Therapy (ACT)~\cite{ACT}, a mindfulness-based cognitive approach, promotes psychological flexibility by encouraging acceptance and value-driven actions, applicable to conditions like chronic pain and anxiety.

Recent advancements in AI have enhanced emotional support systems, addressing limitations of smartphone-based conversational agents~\cite{smartphone}, Muffin framework~\cite{muffin} uses model-agnostic AI feedback and contrastive learning to improve response fluency and relevance. Hybrid Empathetic Framework (HEF)~\cite{HEF} integrates LLMs with small-scale empathetic models to enhance emotion detection and response generation. The Sequential SMES framework~\cite{smes} leverages multimodal data to simulate therapeutic empathy and deliver tailored responses. Our \textsc{MultiMood} framework advances these efforts by incorporating trustworthiness through reinforcement learning with PPO and GRPO, ensuring safe, empathetic, and contextually appropriate responses for diverse user needs.

\subsubsection{Multimodal LLMs} 
Multimodal LLMs like InternVideo2.5~\cite{internvideo2.5} and VideoLLaVA~\cite{videollava} advance video understanding. InternVideo2.5 employs a single InternVIT encoder with Hierarchical Token Compression (HICO) to efficiently process long videos, merging similar tokens to reduce computation while preserving quality~\cite{internvideo2.5}. Its three-stage training supports tasks like temporal grounding and object tracking. VideoLLaVA aligns images and videos using LanguageBind for unified visual representation via a shared projection layer~\cite{videollava}, but lacks audio processing, unlike \textsc{MultiMood}’s modality-specific projectors. These models focus on visual content while missing audio cues (volume, tone, pitch) critical for emotion recognition. Multimodal LLMs also enhance emotional support, overcoming single-modality limitations by capturing nuanced emotional signals for empathetic AI. EmotionLLaMA~\cite{emotionllama} uses the MERR dataset (28,618 samples) and specialized encoders for precise emotion recognition. The SMES framework, with the MESC dataset (28,762 utterances from \textit{In Treatment}), processes multimodal inputs for emotion recognition, strategy prediction, and response generation, improving therapeutic mimicry~\cite{smes}. \textsc{MultiMood} stands out with specialized encoders per modality and a reinforcement learning algorithm designed for trustworthiness, as summarized in Table 1.

\subsubsection{Trustworthiness in Responses}
Trustworthiness is essential for effective emotional support from therapists and doctors, fostering a safe space for patient vulnerability. Goleman’s emotional intelligence framework~\cite{goleman} emphasizes empathy, self-regulation, and social skills as key to building trust, enabling clinicians to communicate effectively. Crits-Christoph et al.~\cite{client-clinician} highlight that trust, distinct from therapeutic alliance, encourages sharing private information, with racial disparities (e.g., lower trust among Black patients) underscoring equity’s role. Richmond et al.~\cite{trustindoctor} link trustworthiness to communication, fidelity, and fairness, noting that lower trust can delay care. In LLMs, \textit{trustworthiness} is critical for safe, supportive interactions, as outlined in TrustLLM’s eight dimensions~\cite{trustllm}: \textit{truthfulness} ensures accuracy, \textit{safety} fosters healthy dialogue, \textit{fairness} promotes impartiality, and \textit{robustness} ensures reliability. \textit{Privacy} protects autonomy, \textit{machine ethics} ensures moral behavior, \textit{transparency} provides clarity, and \textit{accountability} holds LLMs responsible. In \textsc{MultiMood}, these factors are integrated to train robust LLMs, significantly reducing hallucination. Building on these foundations, we propose a tailored set of trustworthiness dimensions for emotional support systems to improve automatically generated responses to meet therapeutic standards.

\section{Methodology}
\subsection{Overview}
\begin{figure*}[htbp]
    \centering
    \includegraphics[width=0.78\linewidth]{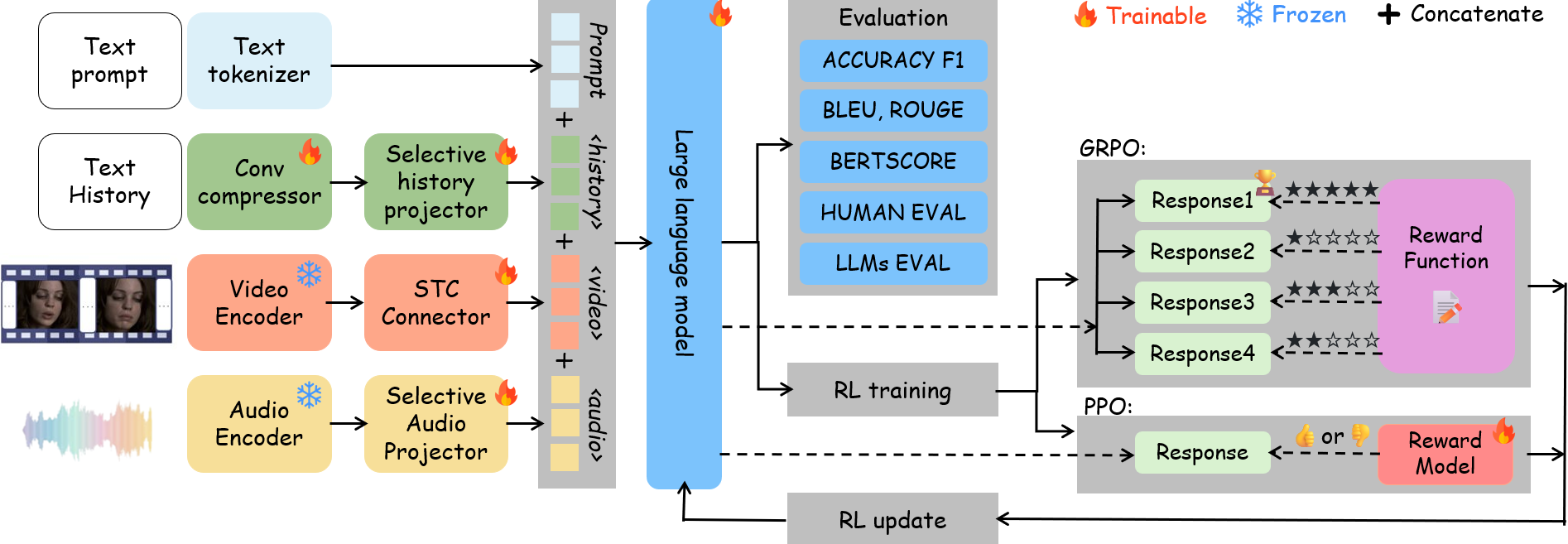}
    \caption{\textsc{MultiMood} overview. Multimodal architecture that processes video, audio, text, and historical conversation data through dedicated encoders. The modality-specific embeddings are fused and passed into an LLM, which is further optimized using reinforcement learning guided by trustworthiness criteria to generate emotionally supportive and responsible responses.}
    \label{fig:overview}
\end{figure*}

The \textsc{MultiMood} framework, shown in Figure \ref{fig:overview}, integrates an audio encoder \(\mathcal{E}^{aud}\), a visual encoder \(\mathcal{E}^{vis}\), a conversation compressor \(\mathcal{C}\), and a large language model \(\phi\). For an input tuple \(P = \langle \text{Audio}, \text{Video}, \text{Prompt}, \text{History} \rangle\), the model is defined as:
\begin{equation} 
    \small
    \hat{O} = \Psi(\phi, \mathcal{E}, \Omega, \mathcal{C}, P),
\end{equation}
where \(\mathcal{E}\) combines audio, vision, and text encoders, \(\Omega\) is the vision pre-processor, and \(\hat{O}\) is the text output. A multi-tower architecture generates modality-specific embeddings: video via vision tower \(f_V\), audio via audio tower \(f_A\), and text via \(\phi\) tokenizer \(f_T\), yielding \(E_T = f_T(\text{Prompt})\). A compressor distills text histories into concise representations, \(E_H = \text{ConvCompressor}(H)\), enabling efficient context processing. Embeddings \([E'_V; E'_A; E_T; E'_H]\) are aligned via modality-specific projectors and fed into the LLM to predict outcomes for four tasks.

\subsection{Framework Components} 

\subsubsection{\textbf{Modality-Specific Encoder}}



The vision pre-processor \(\Omega\) uses the input \(\text{video}\) as a frame sequence, processed by a CLIP-based \cite{CLIP} visual encoder \(\mathcal{E}^{vis}\) to extract video features:
\begin{equation} \small
E_V = \mathcal{E}^{vis}(\Omega(\text{Video})).
\end{equation}
A Spatial-Temporal Convolution (STC) connector \cite{videollama2} captures spatial and temporal dynamics:
\begin{equation} \small
E'_V = \text{STC}(E_V) = P_V(\text{R}_2(\text{Conv3D}(\text{R}_1(E_V)))),
\end{equation}
where \(\text{STC}(\cdot)\) includes two spatial interaction modules (\(\text{R}_1\), \(\text{R}_2\)) and a 3D convolution (\(\text{Conv3D}\)), with \(P_V\) projecting features to the language model \(\phi\) space.

For audio, the BEATs model~\cite{beats} serves as the audio encoder \(\mathcal{E}^{aud}\), extracting features mapped to the language model space via a linear projector \(P_A\):
\begin{equation} 
\small
E_A = \mathcal{E}^{aud}(\text{Audio}), \quad E'_A = P_A(E_A).
\end{equation}

\subsubsection{Cross-Modality Concatenation}
Our approach draws from ECoT~\cite{ECoT}, a plug-and-play method that boosts LLM performance in emotional generation tasks by aligning with Goleman’s emotional intelligence theory. We design a prompt template to guide the LLM in generating empathetic responses, integrating historical and real-time data. Multimodal features are concatenated into the input using specialized tokens: \(\texttt{<video>}\), \(\texttt{<audio>}\), and \(\texttt{<history>}\), replaced by processed embeddings \(E'_V\), \(E'_A\), and \(E'_H\), respectively, forming the input sequence \(X_{LLM}\). This attention-based fusion enables the model to dynamically prioritize relevant cues (e.g., tone, facial expressions) for safe, context-aware responses, while simultaneously predicting three classifications and generating therapist-like outputs (see Figure \ref{fig:overview}). More specific concatenation is discussed in the Appendix A.1.

\subsubsection{\textbf{Conversation Compressor}}

\begin{figure}[htbp]
\includegraphics[width=1.0\linewidth]{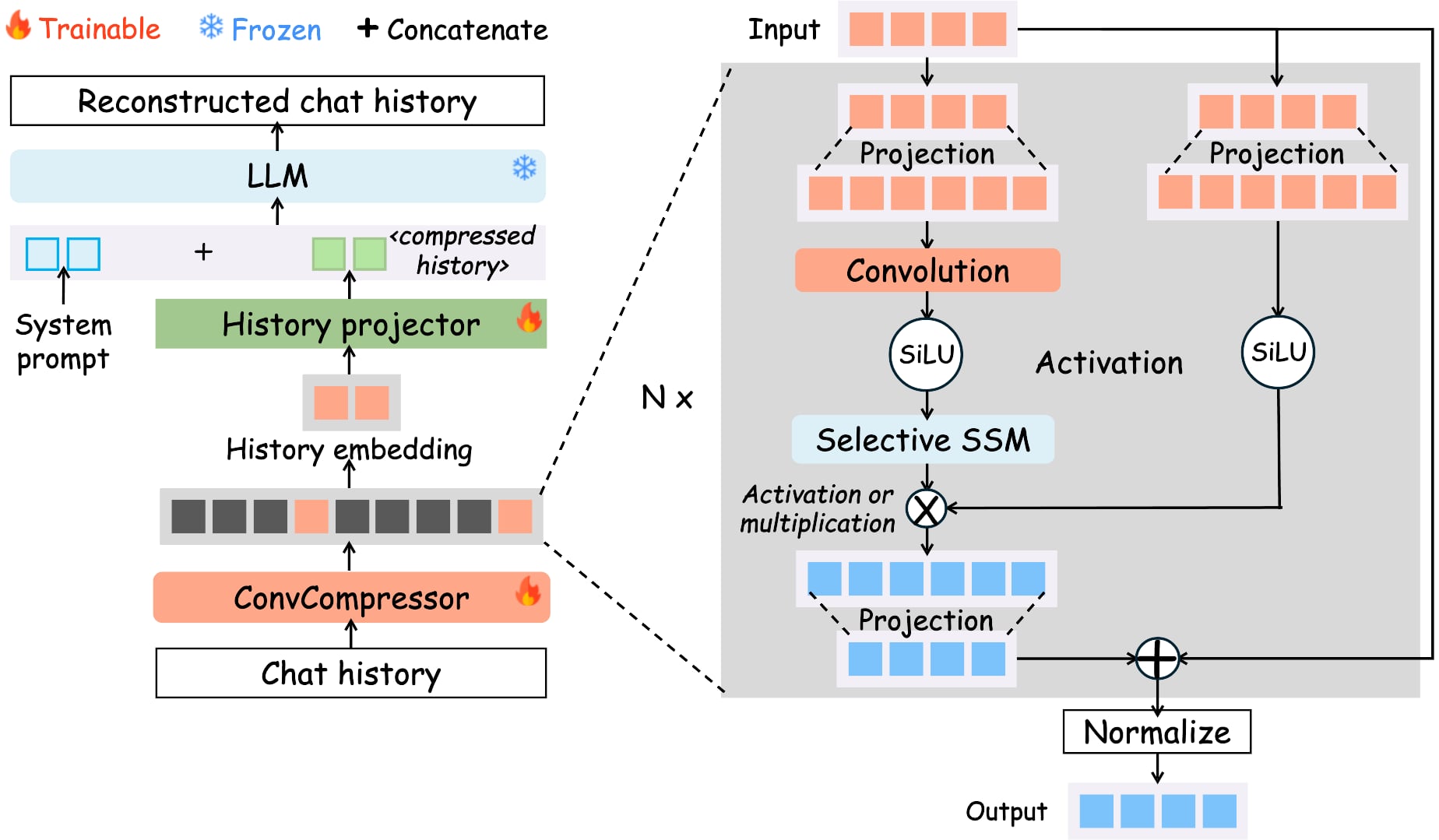}
    \caption{ConvCompressor architecture and pretraining.}
    \label{fig:convcom}
\end{figure}
Conversational emotional support systems require effective processing of extensive dialogue histories to deliver contextually appropriate responses. However, long conversation histories pose computational and memory challenges for language models. To address this, we propose Conversation Compressor (ConvCompressor), a lightweight module that distills dialogue histories into compact, semantically rich representations while retaining critical information. ConvCompressor employs the Mamba state-space model~\cite{mamba} as its core, offering linear computational complexity compared to the quadratic scaling of transformers. It appends a \texttt{<MEM>} token to each conversational turn $U_i$ in a history $H = \text{concat}(\{U_{i}\}_{i=1}^{T})$, where $U_i$ includes role information, utterance content, emotional labels, and therapist strategy labels, forming $H'=U_1\texttt{<MEM>}U_2\texttt{<MEM>}\ldots U_T\texttt{<MEM>}$. The Mamba backbone processes $H'$ to generate hidden representations $Z$, from which we extract hidden states only at \texttt{<MEM>} token positions. The extracted representations then undergo a trainable memory projector $P_H$ before being fed to LLM (see Figure~\ref{fig:convcom}).

ConvCompressor is optimized through a two-stage training process. First, it is pre-trained with a frozen language model on a reconstruction task to regenerate the original dialogue history from compressed \texttt{<MEM>} representations. Then, it undergoes end-to-end fine-tuning within the multimodal pipeline, adapting its compression strategy to significantly reduce the number of input tokens for the LLM while preserving a comparative overall performance.

\begin{table*}[]

\begin{minipage}{0.6\textwidth}
\centering
\textbf{Trustworthiness Dimensions for Emotional Support}\\
\scalebox{0.75}{
\setlength{\tabcolsep}{3pt}
\begin{tabular}{l l p{5.5cm}}
\toprule
\textbf{Dimension} & \textbf{Source} & \textbf{Definition} \\
\hline
\textbf{Truthfulness} & \begin{tabular}[t]{@{}l@{}} \cite{trustllm} \\ \cite{trustindoctor} \\ \cite{goleman} \end{tabular} & The accurate representation of information, facts, and results by the AI system. \\
\hline
\textbf{Safety} & \begin{tabular}[t]{@{}l@{}} \cite{trustllm} \\ \cite{goleman} \end{tabular} & Promote safe, healthy conversations, avoiding harm, distress, or triggers while supporting user well-being. \\
\hline
\textbf{Fairness} & \begin{tabular}[t]{@{}l@{}} \cite{trustllm} \\ \cite{trustindoctor} \\ \cite{goleman} \end{tabular} & The quality of being impartial and equitable, considering multiple perspectives and maintaining a positive, action-oriented tone. \\
\hline
\textbf{Privacy} & \begin{tabular}[t]{@{}l@{}} \cite{trustllm} \\ \cite{trustindoctor} \end{tabular} & Practices that safeguard human autonomy, identity, and data dignity. \\
\hline
\textbf{Empathy} & \begin{tabular}[t]{@{}l@{}} \cite{trustindoctor} \\ \cite{goleman} \end{tabular} & Openness and honesty in expressing sympathy for negative situations or approval for positive ones. \\
\hline
\textbf{Reliability} & \begin{tabular}[t]{@{}l@{}} \cite{client-clinician} \\ \cite{trustindoctor} \\ \cite{goleman} \end{tabular} & Responses foster understanding, connection, and provide encouragement, comfort, or support. \\
\hline
\textbf{Ethical Guidance} & \begin{tabular}[t]{@{}l@{}} \cite{trustllm} \\ \cite{goleman} \end{tabular} & Ensuring AI behaviors guide emotional health responsibly, avoiding manipulation or harm. \\
\bottomrule
\end{tabular}}
\end{minipage}
\hfill
\begin{minipage}{0.38\textwidth}
\centering
\scalebox{0.8}{
\setlength{\tabcolsep}{2pt}
\begin{tabular}{lccccc}
\multicolumn{5}{c}{\textbf{Inter-annotator Agreement (Top)}} \\
\textbf{Dimension}    & \textbf{Flu.} & \textbf{Ide.} & \textbf{Com.} & \textbf{Sug.} & \textbf{Ove.} \\ \hline
\textbf{Fleiss Kappa} & 0.65          & 0.61          & 0.60          & 0.61          & 0.67       \\
[0.3em]
\multicolumn{5}{c}{\textbf{Human Evaluation (Middle)}} \\
                            & \textbf{Flu.} & \textbf{Ide.} & \textbf{Com.} & \textbf{Sug.} & \textbf{Ove.} \\ \hline
\textbf{Qwen2-7B}          & 22\%          & 17\%          & 17\%          & 22\%          & 21\%          \\
\textbf{MM (SFT)}    & 23\%          & 30\%          & 27\%          & 25\%          & 21\%          \\
\textbf{MM (SFT+RL)} & \textbf{55\%} & \textbf{53\%} & \textbf{56\%} & \textbf{53\%} & \textbf{58\%} \\
\end{tabular}}

\scalebox{0.8}{
\setlength{\tabcolsep}{0.8pt}
\begin{tabular}{lcccccccc}
\multicolumn{9}{c}{\textbf{LLMs Evaluation (Bottom)}} \\[0.2em]
\multicolumn{1}{c}{\multirow{2}{*}{\textbf{Model}}} & \multicolumn{8}{c}{\textbf{Judge: GPT-4o}}                   \\ \cline{2-9} 
\multicolumn{1}{c}{}                          & Tru. & Saf. & Fai. & Pri. & Emp. & Rel. & Eth. & Avg. \\ \hline
\textbf{Qwen2-7B} & 6.0 & 4.2 & 5.1 & 8.0 & 4.2 & 4.6 & 4.3 & 5.2 \\
\textbf{MM (SFT)} & 6.2 & 4.3 & 5.8 & 7.8 & 4.3 & 4.8 & 4.9 & 5.4 \\
\textbf{MM (SFT+RL)} & \textbf{7.0} & \textbf{6.3} & \textbf{6.8} & \textbf{8.8} & \textbf{6.2} & \textbf{6.4} & \textbf{6.3} & \textbf{6.8} \\ \hline
                                              & \multicolumn{8}{c}{\textbf{Judge: Claude 4.0-Sonnet}}        \\ \hline
\textbf{Qwen2-7B}                               & 4.9    & 7.0    & 6.0    & \textbf{8.0}    & 5.6    & \textbf{7.0}    & 6.0    & 6.4    \\
\textbf{MM (SFT)}            & 5.0    & \textbf{7.0}   & 5.9    & \textbf{8.0}    & 5.9    & 7.3    & 6.0        & 6.5    \\
\textbf{MM (SFT+RL)}                   & \textbf{7.2}    & \textbf{7.0}    & \textbf{6.5}    & \textbf{8.0}    & \textbf{7.8}    & \textbf{7.4}    & \textbf{6.6}    & \textbf{7.2}    \\ \hline
                                              & \multicolumn{8}{c}{\textbf{Judge: Grok-3}}                   \\ \hline
\textbf{Qwen2-7B} & 6.1 & 6.0 & 7.2 & 7.5 & 6.0 & 6.3 & 5.8 & 6.4 \\
\textbf{MM (SFT)} & 6.2 & 6.6 & 7.6 & 7.2 & 6.1 & 6.9 & 5.9 & 6.6 \\
        \textbf{MM (SFT+RL)} & \textbf{7.3} & \textbf{7.8} & \textbf{8.7} & \textbf{9.3} & \textbf{7.5} & \textbf{7.9} & \textbf{7.5} & \textbf{8.0} \\
\end{tabular}}
\end{minipage}
\caption{Trustworthiness dimensions for emotional support tasks (left); Inter-annotator agreement, human and evaluation results (right) across different models. Flu., Ide., Com., Sug., Ove., stand for Fluency, Identification, Comfort, Suggestions, and Overall, whose definitions are provided in the Appendix C.1.}
\label{tab:trustworthiness_evaluation}
\end{table*}

\subsection{Training}
\label{sec:training}
The training process comprises two key stages that enhance both robustness and trustworthiness. 
 
\noindent\textbf{Stage 1: Supervised Fine-Tuning} We fine-tune our framework on the MESC dataset, leveraging multimodal data throughout the training process. To accommodate potential missing modalities (video or audio) during inference, we introduce a random modal selection mechanism. This is defined by a probability vector \(\mathbf{p} = [p_a, p_v, p_{av}]\), representing the likelihoods of selecting audio or video or both modalities. This approach enhances the framework's robustness by exposing it to all possible modality combinations during training. For multimodal processing, we employed SigLIP-So400M-Patch14 384~\cite{DBLP:siglip/iccv/ZhaiM0B23} for video, and BEATs~\cite{beats} for audio. The ConvCompressor is built on Mamba-370M~\cite{mamba}. Prompt for training is provided in the Appendix A.1.

\noindent\textbf{Stage 2: Trustworthiness-Aware via Reinforcement Learning}  
Initial assessments showed that the system's responses occasionally lacked the natural flow and trustworthiness needed for effective emotional support. To address this, we defined a set of trustworthiness criteria, \(\mathcal{C} = \{c_1, c_2, \dots, c_k\}\) (see Table~\ref{tab:trustworthiness_evaluation}-left), and employed reinforcement learning to align responses with ethical and therapeutic standards. We used Group Relative Policy Optimization (GRPO)~\cite{grpo} and Proximal Policy Optimization (PPO)~\cite{ppo} after supervised fine-tuning to enhance response quality. GRPO optimizes by comparing responses within groups, while PPO stabilizes learning through clipped updates and a Kullback-Leibler divergence penalty. To guide learning, we designed a reward function combining trustworthiness and similarity. The similarity score \( r_{\text{sim}}(y, y^*) \) leverages BGE-M3 embeddings via ColBERT~\cite{Colbert}, integrating dense, sparse, and ColBERT-specific similarities with weights (1, 0.3, 1), normalized to \([0, 1]\):  
\begin{equation}
\small
r_{\text{sim}} = \text{scale}\left( \text{sim}_{\text{C}} + 0.3\, \text{sim}_{\text{s}} + \text{sim}_{\text{d}} \right)
\end{equation}
Trustworthiness \( r_{\text{trust}}(y) \) is evaluated by GPT-4o per sentence, averaged and scaled to \([0, 1]\). GPT-4o is trusted for this task due to its proven capabilities in labeling data~\cite{llm-anno} across various tasks, as well as its robust safety mechanisms ~\cite{safety-mechanisms}:  
\begin{equation}
\small
r_{\text{trust}}(y) = \text{scale}\left(\frac{1}{|y|}\sum_{i=1}^{|y|} \text{GPT-4o}_{\text{trust}}(y_i)\right)
\end{equation}
The final reward is:  
\begin{equation}
\small
r(y|x) = \frac{1}{2} \left( r_{\text{trust}}(y) + r_{\text{sim}}(y, y^*)\right)
\label{eq:reward}
\end{equation}

Details of RL strategies are in the Appendix A.2.

\begin{table*}[ht]
\centering
\small
\begin{tabular}{lccccccccc}
\hline
\multicolumn{1}{c}{\textbf{Method}} & \textbf{Hap} & \textbf{Sad} & \textbf{Neu} & \textbf{Ang} & \textbf{Sur} & \textbf{Dis} & \textbf{Fea} & \textbf{UAR} & \textbf{WAR} \\ \hline
IAL \cite{IAL} & 87.95 & 67.21 & 70.10 & 76.06 & 62.22 & 0.00 & 26.44 & 55.71 & 69.24 \\ 
VideoMAE \cite{videoMae} & 93.09 & 78.78 & 71.75 & 78.74 & 33.44 & 17.93 & 41.46 & 63.60 & 74.60 \\ 
S2D \cite{S2D} & 93.62 & 80.25 & 77.14 & 81.09 & 64.53 & 1.38 & 34.71 & 61.82 & 76.03 \\ 
EmotionLLaMA \cite{emotionllama} & 93.05 & 79.42 & 72.47 & 84.14 & 72.79 & 3.45 & 44.20 & 64.21 & 77.06 \\ 
MultiMood (ours) & \textbf{96.31} & \textbf{93.68} & \textbf{89.45} & \textbf{88.82} & \textbf{81.68}  & \textbf{78.38} & \textbf{85.19} & \textbf{85.94} & \textbf{89.89} \\ \hline
\end{tabular}
\caption{Comparison of multimodal emotion recognition results on DFEW.}
\label{tab:DFEW}
\end{table*}

\subsection{Trustworthiness Dimension Table} \label{sec:dimension-table}

To assess response trustworthiness in emotional support, we first developed a domain-specific framework, \textit{Trustworthy Dimensions}. This was built by synthesizing insights from four key sources: the TrustLLM framework~\cite{trustllm}, which outlines trust principles for LLMs; patient-clinician trust studies~\cite{client-clinician, trustindoctor}; and Goleman’s emotional intelligence principles~\cite{goleman}. From TrustLLM, we adopted core technical values such as Truthfulness, Safety, Fairness, Privacy, and Machine Ethics. Clinical trust literature contributed Honesty, Communication, Confidentiality, Fidelity, and Reliability - emphasizing relational trust. Goleman’s work added Empathy and Social Skills, highlighting emotional resonance. These elements were distilled into seven core dimensions, carefully defined to balance technical reliability with emotional sensitivity (Table~\ref{tab:trustworthiness_evaluation}-left). Details of each criterion, informed by prior literature, are provided in the Appendix A.3.

\section{Experiments}

\subsection{Experimental Setup}
\subsubsection{Metrics}
For generation evaluation, we use BLEU-n (B-2), ROUGE-L (R-L), and BERTScore (BS) to evaluate the Therapist's responses from the models. For classification of the MESC dataset \cite{smes}, we use Accuracy and Weighted-F1 as metrics. These metrics collectively provide a comprehensive overview of model performance across different tasks. For the DFEW dataset \cite{DFEW}, we use unweighted average recall (UAR) and weighted average recall (WAR) to compare our method with SOTA methods.

\subsubsection{Baselines}\label{part:llmbaseline}
We utilized the pretrained LLM from VideoLLaMA2~\cite{videollama2} as the multimodal LLM backbone, leveraging its training on multimodal data. We compare \textsc{MultiMood} with API-based LLMs (GPT-4o~\cite{GPT4}, Grok3~\cite{xai2025grok2}, Claude-3.7~\cite{claude}, Deepseek-R1~\cite{deepseek}, LLaMA4~\cite{llama4}); Open-source VLMs (Qwen2~\cite{qwen2} and Qwen2.5~\cite{qwen25}, EmotionLLaMA~\cite{emotionllama}, VideoLLaMA2, VideoLLaVA~\cite{lin-etal-2024-video-llava}, InternVideo2.5-8B~\cite{internvideo2.5}, VideoLLaMa3-7B~\cite{videollama3}) - finetuned on the MESC dataset~\cite{smes} with SFT and PPO; VideoLLaMA2-72B and closed sources models are not finetuned due to resource constraints; and SMES-leveraged models~\cite{smes}. 
\subsubsection{Settings} Experiments were conducted on 4$\times$H100 GPUs including LLMs training, multimodal projectors training, ConvCompressor training and RL training. 
Detailed experiment setup information are discussed in Appendix B.1.


\subsection{Results}
In this section, we present a comprehensive evaluation to compare our framework with other frontier models on the MESC and DFEW datasets. The evaluation highlights the strengths and advancements of our framework in handling complex multimodal data in both tasks. Ablation study of modalities affect show in Appendix D.

\begin{table*}[htbp]
\centering
\setlength{\tabcolsep}{4pt}
\footnotesize
\scalebox{1}{
\begin{tabular}{l c c ccccccccc}
\hline
\textbf{Model} & \textbf{Training method} & \textbf{Modality} &
\multicolumn{2}{c}{\textbf{Task 1}} &
\multicolumn{2}{c}{\textbf{Task 2}} &
\multicolumn{2}{c}{\textbf{Task 3}} &
\multicolumn{3}{c}{\textbf{Task 4}} \\
\cline{4-12}
& & & Acc & F1 & Acc & F1 & Acc & F1 & B2 & R-L & BScore \\
\hline
MMGCN & SFT & A,V,T & 55.80 & 57.58 & - & - & - & - & - & - & - \\
MMDFN & SFT & A,V,T & 58.13 & 55.86 & - & - & - & - & - & - & - \\
Blenderbot SFT & SFT & A,V,T & - & - & - & - & 48.00 & \textbf{46.10} & 1.31 & 15.38 & 86.60 \\
SMES & SFT & A,V,T & 54.60 & 46.80 & 96.10 & 64.00 & \textbf{49.00} & 20.20 & 5.13 & 15.42 & \textbf{86.80} \\
\hline
VideoLLaMA2-72B & - & A,V,T & 55.06 & 55.68 & 97.36 & 98.10 & 25.77 & 26.09 & 3.55 & 13.77 & 85.37 \\
VideoLLaVA & SFT & V,T & 46.60 & 47.08 & 94.18 & 88.03 & 27.31 & 22.28 & 4.37 & 9.84 & 84.23 \\
InternVideo2.5-8B & SFT & V,T & 37.22 & 34.69 & 98.90 & 98.79 & 13.44 & 4.82 & 3.92 & 13.21 & 85.40 \\
VideoLLaMA3-7B & SFT & A,V,T & 45.28 & 46.23 & 97.40 & 72.66 & 33.96 & 24.50 & 3.70 & 11.55 & 85.07 \\
EmotionLLaMA & SFT & A,V,T & 46.12 & 41.95 & \textbf{99.11} & \textbf{99.11} & 37.44 & 25.41 & 2.55 & 10.76 & 84.28 \\
\hline
LLaMA4-Maverick & - & T & 23.34 & 21.16 & 68.72 & 81.02 & 14.53 & 8.11 & 3.94 & 10.03 & 84.26 \\
Claude-3.7-Sonnet & - & T & 32.59 & 33.33 & 85.90 & 91.80 & 27.97 & 27.55 & 2.25 & 8.45 & 83.79 \\
Deepseek-R1 & - & T & 20.48 & 20.27 & 59.47 & 74.03 & 17.84 & 15.95 & 3.22 & 9.20 & 83.96 \\
GPT-4o & - & T & 38.98 & 43.56 & 72.46 & 83.60 & 24.88 & 26.26 & 2.30 & 9.20 & 84.31 \\
Grok-2 & - & T & 22.46 & 25.08 & 65.85 & 78.80 & 20.44 & 18.19 & 2.29 & 9.60 & 84.61 \\
\hline
Qwen2-7B & SFT & A,V,T & 41.83 & 37.16 & \textbf{99.33} & 99.00 & 37.43 & 33.52 & 4.68 & 13.20 & 85.61 \\
Qwen2-0.5B & SFT+PPO & A,V,T & 44.27 & 44.99 & \textbf{99.33} & 99.00 & 36.34 & 33.01 & 4.40 & 12.31 & 85.36 \\
Qwen2-7B  & SFT+Comp. & A,V,T & 44.27 & 44.03 & \textbf{99.33} & 99.00 & 39.42 & 35.69 & 4.60 & 12.90 & 85.47 \\
Qwen2.5-7B & SFT & A,V,T & 53.00 & 51.13 & 98.63 & 98.80 & 35.14 & 34.46 & 4.68 & 13.81 & 85.52 \\
\hline
\textbf{MultiMood } & \textbf{SFT+Comp.} & A,V,T & 53.75 & 51.75 & \textbf{99.33} & 99.00 & 39.29 & 36.25 & 5.26 & 15.34 & 85.81 \\
\textbf{MultiMood } & \textbf{SFT} & A,V,T & 56.38 & 55.81 & 99.11 & \textbf{99.11} & 36.78 & 34.32 & 4.58 & 13.47 & 85.71 \\
\textbf{MultiMood } & \textbf{SFT+GRPO} & A,V,T & \textbf{58.60} & \textbf{57.78} & \textbf{99.33} & 99.00 & 42.81 & 39.65 & \textbf{6.18} & \textbf{17.86} & \textbf{86.80} \\
\textbf{MultiMood } & \textbf{SFT+Comp+GRPO} & A,V,T & 55.94 & 55.33 & 99.11 & \textbf{99.11} & 38.10 & 34.58 & 5.42 & 15.83 & 86.00 \\
\hline
\end{tabular}}
\caption{Benchmark of \textsc{MultiMood} against other baselines on MESC. Task 1: User Emotion Recognition, Task 2: System Emotion Recognition, Task 3: Strategy Prediction, Task 4: Response Generation. A=Audio, V=Video, T=Text; B2=BLEU-2; R-L=ROUGE-L; BScore=BERTScore (F1); Comp.=Conversation Compressor.}
\label{tab:main_result}
\end{table*}

\subsubsection{\textbf{Overall Performance}}
\noindent Tables \ref{tab:DFEW} and \ref{tab:main_result} present the primary results of our proposed MultiMood framework compared to baseline models, evaluated across four MESC tasks \cite{smes} and one DFEW task \cite{DFEW}. MultiMood demonstrates exceptional adaptability, achieving robust performance across all tasks, unlike baseline models that often excel in specific domains. It delivers consistent results in emotion recognition, strategy prediction, system emotion prediction, and response generation, surpassing larger models like VideoLLaMA2-72B and specialized classification models like MMGCN. Notably, MultiMood achieves the highest average score (56.45) across the four MESC tasks and a SOTA score on the DFEW dataset. The ConvCompressor module demonstrates remarkable efficiency, achieving 98.6\% token reduction (see result in Appendix B.2)
while maintaining competitive performance, making our framework significantly more memory-efficient for processing extended dialogue histories. Our framework performance is evaluated from four key perspectives.

\noindent\texttt{\textbf{Emotion Recognition:}} Our \textsc{MultiMood} framework achieves SOTA performance on the single-labeled DFEW dataset \cite{DFEW}, outperforming prior methods in accuracy, unweighted average recall and weighted average recall scores, as shown in Table \ref{tab:DFEW}. It achieves the highest UAR of 85.94\% and WAR of 89.89\%, excelling across all emotion categories, notably Disgust (78.38\%), where prior models like IAL \cite{IAL}, VideoMAE \cite{videoMae}, S2D \cite{S2D}, and EmotionLLaMA \cite{emotionllama} struggled due to under-representation \cite{DFEW}. With MESC, the variant utilizing GRPO attains the highest performance, followed closely by the fine-tuned framework without GRPO. \textsc{MultiMood} surpasses video understanding models (e.g., VideoLLaMA, InternVideo2.5), the Qwen family, and closed-source models, as well as specialized frameworks like SMES \cite{smes} and MMDFN \cite{MMDFN} (shown in Table~\ref{tab:main_result}).  \textsc{MultiMood}'s robust classification, particularly for nuanced emotions, enhances empathetic response generation, establishing a new benchmark for precise emotion recognition. However, while ConvCompressor improves memory efficiency, it may compromise performance due to information loss.

\noindent \texttt{\textbf{Strategy Prediction:}} \textsc{MultiMood} achieves a 42.81\% accuracy on the Strategy Prediction task, slightly trailing BlenderBot SFT (48\%) and SMES (49\%) \cite{smes}. This gap reflects \textsc{MultiMood}'s design prioritizing robust, generalized performance across diverse tasks over specialization in strategy prediction. Nonetheless, it delivers a competitive F1 score, surpassing several baselines, though marginally behind SMES in accuracy. Unlike BlenderBot, which benefits from domain-specific retrieval tools, \textsc{MultiMood} faces challenges with class imbalance. However, its instruction-guided framework excels in generating safe, multimodal-aware responses, enhancing generalizability across emotion recognition, strategy planning, and empathetic response generation.

\noindent\texttt{\textbf{System Emotion Prediction:}} Most fine-tuned models achieve over 90\% accuracy in this task, attributed to a data skew where 90\% of labels are Neutral. This imbalance is typical in emotional support contexts, as therapists maintain a calm demeanor, enables the system to generate honest, unbiased answers.

\noindent\texttt{\textbf{Response Generation:}} 
\textsc{MultiMood} (SFT+GRPO) achieves superior performance across all metrics--BLEU-2 (6.18), ROUGE-L (17.86), and BERTScore (86.80)--demonstrating the efficacy of combining Group Relative Policy Optimization with supervised fine-tuning to produce fluent, contextually aligned responses. It outperforms baselines like VideoLLaMA2-7B (SFT) and Qwen2-7B (SFT + PPO), as well as closed-source models such as GPT-4o \cite{GPT4} and LLaMA4 \cite{llama4}, which underperform due to their reliance on textual features alone. \textsc{MultiMood}'s integration of multimodal data enhances its classification and response generation capabilities, surpassing recent SOTA SMES \cite{smes} and setting a new benchmark for empathetic, high-quality responses. Some examples of response show in Appendix E.


\subsubsection{\textbf{Human and LLM Evaluation}}

We conducted a comprehensive evaluation using both human and LLM assessments to assess the trustworthiness and quality of responses from Qwen2-7B (SFT), MultiMood--MM(SFT), and MultiMood--MM(SFT+RL). Four graduate students served as human annotators, all with expertise in emotional support research and advanced English proficiency (IELTS overall $\geq$ 7.0 with reading $\geq$ 7.5) to ensure accurate evaluation of text-only outputs. They received training with tutorials and examples, including framework-generated outputs, dialogue contexts, situational details, and responses from a licensed psychologist, followed by a test on 100 MESC dataset validation samples to achieve a Cohen’s kappa inter-annotator agreement above 0.4 \cite{cohenkappa} (see Table \ref{tab:trustworthiness_evaluation}-right-middle); retraining was required if unmet. During annotation, two annotators labeled all responses, with discrepancies resolved by a third and persistent disagreements settled by a fourth to establish the majority label, detailed results in Table \ref{tab:trustworthiness_evaluation}-right-middle. The annotation guideline is provided in the Appendix C.1.

Human evaluation shows MultiMood (SFT+GRPO) outperforming in Fluency (55\%), Comfort (56\%), and Overall (58\%), highlighting the effectiveness of multimodal fine-tuning and GRPO in enhancing response quality. Simultaneously, LLM evaluation, guided by \cite{llm-anno}, \cite{llm-medical-1}, and \cite{llm-medical-2}, underscored LLMs’ near-human accuracy in surgical knowledge but noted query inconsistency, stressing stable evaluation needs. LLM scoring pre- and post-application of our trustworthiness dimension table \ref{tab:trustworthiness_evaluation}-left revealed RL-incorporated frameworks significantly outperformed non-RL frameworks across three LLMs (see Table \ref{tab:trustworthiness_evaluation}-right-bottom). By aligning with trustworthiness criteria, RL enhances safety, reliability, and ethical soundness, addressing non-RL inconsistencies and boosting utility for critical applications. Prompt for LLMs evaluation is provided in the Appendix C.2. 

\section{Limitation}

Despite the demonstrated effectiveness of our framework, several limitations persist. It underperforms BlenderBot in strategy prediction (per SMES~\cite{smes}) due to class imbalance and the lack of external retrieval. The inability to fine-tune certain multimodal frameworks, constrained by resource limitations, weakens the robustness of our comparisons. Additionally, although its usage was proved~\cite{llm-anno}, using GPT-4o for trustworthiness evaluation may introduce bias, particularly when its reward function influences reinforcement learning training. Furthermore, the experimental datasets, derived from movies and challenges rather than real treatment settings, lack authenticity—a common issue in this field~\cite{medical-data-1, medical-data-2}, underscoring the need for more realistic emotion support datasets in future research.

\section{Conclusion}

In conclusion, \textsc{MultiMood} leverages multimodal techniques to achieve state-of-the-art results in emotion recognition and response generation, outperforming closed- and open-source models. Enhanced by reinforcement learning, it demonstrates high trustworthiness per human and LLM evaluations, with potential for therapeutic use. However, limitations in strategy prediction, hardware constraints and lack of realistic datasets suggest areas for future enhancement.

\section{Acknowledgments}
This research was supported by the Mohamed bin Zayed University of Artificial Intelligence Travel Grant. The authors also gratefully acknowledge the VNUHCM-University of Information Technology’s Scientific Research Support Fund for their financial assistance.  The
authors also thank the International Max Planck Research School for Intelligent Systems (IMPRS-IS)
for supporting Duy M. H. Nguyen. Duy M. H. Nguyen and Daniel Sonntag are also supported by the No-IDLE project (BMFTR, 16IW23002), the MASTER project
(EU, 101093079), and the Endowed Chair of Applied Artificial Intelligence, Oldenburg University.



\bibliography{aaai2026}

@ARTICLE{smes,
  author={Chu, Yuqi and Liao, Lizi and Zhou, Zhiyuan and Ngo, Chong-Wah and Hong, Richang},
  journal={IEEE Transactions on Multimedia}, 
  title={Towards Multimodal Emotional Support Conversation Systems}, 
  year={2025},
  volume={},
  number={},
  pages={1-12},
  doi={10.1109/TMM.2025.3604951}}

@book{goleman1995emotional,
  author = {Goleman, Daniel},
  title = {Emotional Intelligence: Why It Can Matter More Than IQ},
  year = {1995},
  publisher = {Bantam Books}
}

@misc{ECoT,
      title={Enhancing Emotional Generation Capability of Large Language Models via Emotional Chain-of-Thought}, 
      author={Zaijing Li and Gongwei Chen and Rui Shao and Yuquan Xie and Dongmei Jiang and Liqiang Nie},
      year={2024},
      eprint={2401.06836},
      archivePrefix={arXiv},
      primaryClass={cs.CL},
      url={https://arxiv.org/abs/2401.06836}, 
}

@misc{WHO,
  author =       "WHO",
  year =         "2022",
  title =        "Mental disorders",
  howpublished =          "\url{https://www.who.int/news-room/fact-sheets/detail/mental-disorders}",
  note="Accessed: 2025-11-11",
}

@misc{NAMI,
  author =  "NAMI",
  year = "2023",
  title = "Mental Health By the Numbers",
  howpublished = "\url{https://www.nami.org/about-mental-illness/mental-health-by-the-numbers/}",
  note="Accessed: 2025-11-11",
}

@article{6billions,
  author = {Marquez, Patricio V. and Saxena, Shekhar},
  title = {Making Mental Health a Global Priority},
  journal = {Cerebrum: The Dana Forum on Brain Science},
  volume = {2016},
  pages = {cer-10-16},
  year = {2016}
}

@inproceedings{IAL,
  author       = {Hanting Li and
                  Hongjing Niu and
                  Zhaoqing Zhu and
                  Feng Zhao},
  editor       = {Brian Williams and
                  Yiling Chen and
                  Jennifer Neville},
  title        = {Intensity-Aware Loss for Dynamic Facial Expression Recognition in
                  the Wild},
  booktitle    = {AAAI 2023},
  pages        = {67--75},
  year         = {2023},
  doi          = {10.1609/AAAI.V37I1.25077},
  bibsource    = {dblp computer science bibliography, https://dblp.org}
}

@inproceedings{videoMae,
  author       = {Zhan Tong and
                  Yibing Song and
                  Jue Wang and
                  Limin Wang},
  editor       = {Sanmi Koyejo and
                  S. Mohamed and
                  A. Agarwal and
                  Danielle Belgrave and
                  K. Cho and
                  A. Oh},
  title        = {VideoMAE: Masked Autoencoders are Data-Efficient Learners for Self-Supervised
                  Video Pre-Training},
  booktitle    = {NeurIPS 2022},
  year         = {2022},
  url          = {http://papers.nips.cc/paper\_files/paper/2022/hash/416f9cb3276121c42eebb86352a4354a-Abstract-Conference.html},
  bibsource    = {dblp computer science bibliography, https://dblp.org}
}

@inproceedings{DBLP:siglip/iccv/ZhaiM0B23,
  author       = {Xiaohua Zhai and
                  Basil Mustafa and
                  Alexander Kolesnikov and
                  Lucas Beyer},
  title        = {Sigmoid Loss for Language Image Pre-Training},
  booktitle    = {{ICCV} 2023},
  pages        = {11941--11952},
  year         = {2023},
  doi          = {10.1109/ICCV51070.2023.01100},
  bibsource    = {dblp computer science bibliography, https://dblp.org}
}

@inproceedings{DFEW,
  author       = {Xingxun Jiang and
                  Yuan Zong and
                  Wenming Zheng and
                  Chuangao Tang and
                  Wanchuang Xia and
                  Cheng Lu and
                  Jiateng Liu},
  editor       = {Chang Wen Chen and
                  Rita Cucchiara and
                  Xian{-}Sheng Hua and
                  Guo{-}Jun Qi and
                  Elisa Ricci and
                  Zhengyou Zhang and
                  Roger Zimmermann},
  title        = {{DFEW:} {A} Large-Scale Database for Recognizing Dynamic Facial Expressions
                  in the Wild},
  booktitle    = {{MM} '20},
  pages        = {2881--2889},
  year         = {2020},
  doi          = {10.1145/3394171.3413620},
  bibsource    = {dblp computer science bibliography, https://dblp.org}
}

@article{S2D,
  author       = {Yin Chen and
                  Jia Li and
                  Shiguang Shan and
                  Meng Wang and
                  Richang Hong},
  title        = {From Static to Dynamic: Adapting Landmark-Aware Image Models for Facial
                  Expression Recognition in Videos},
  journal      = {{IEEE} Trans. Affect. Comput.},
  volume       = {16},
  number       = {2},
  pages        = {624--638},
  year         = {2025},
  doi          = {10.1109/TAFFC.2024.3453443},
  bibsource    = {dblp computer science bibliography, https://dblp.org}
}

@article{qwen25,
  author       = {Shuai Bai},
  title        = {Qwen2.5-VL Technical Report},
  journal      = {CoRR},
  year         = {2025},
  doi          = {10.48550/ARXIV.2502.13923},
  bibsource    = {dblp computer science bibliography, https://dblp.org}
}

@Article{woebot,
author="Fitzpatrick, Kathleen Kara
and Darcy, Alison
and Vierhile, Molly",
title="Delivering Cognitive Behavior Therapy to Young Adults With Symptoms of Depression and Anxiety Using a Fully Automated Conversational Agent (Woebot): A Randomized Controlled Trial",
journal="JMIR Ment Health",
year="2017",
volume="4",
number="2",
pages="e19",
issn="2368-7959",
doi="10.2196/mental.7785",
}

@inproceedings{textonly1,
    title = "Towards Empathetic Open-domain Conversation Models: A New Benchmark and Dataset",
    author = "Rashkin, Hannah  and
      Smith, Eric Michael  and
      Li, Margaret  and
      Boureau, Y-Lan",
    booktitle = "ACL 2019",
    month = jul,
    year = "2019",
    doi = "10.18653/v1/P19-1534",
    pages = "5370--5381",
}

@inproceedings{emotionllama,
  author       = {Zebang Cheng and
                  Zhi{-}Qi Cheng and
                  Jun{-}Yan He and
                  Kai Wang and
                  Yuxiang Lin and
                  Zheng Lian and
                  Xiaojiang Peng and
                  Alexander G. Hauptmann},

  title        = {Emotion-LLaMA: Multimodal Emotion Recognition and Reasoning with Instruction
                  Tuning},
  booktitle    = {NeurIPS 2024},
  year         = {2024},
  url          = {http://papers.nips.cc/paper\_files/paper/2024/hash/c7f43ada17acc234f568dc66da527418-Abstract-Conference.html},
  bibsource    = {dblp computer science bibliography, https://dblp.org}
}

@inproceedings{summarizing,
  author       = {Huy M. Le and
                  Vy T. Luong and
                  Ngoc Hoang Luong},
  title        = {Data Augmentation with Large Language Models for Vietnamese Abstractive
                  Text Summarization},
  booktitle    = {{MAPR} 2023},
  pages        = {1--6},
  year         = {2023},
  doi          = {10.1109/MAPR59823.2023.10288906},
  bibsource    = {dblp computer science bibliography, https://dblp.org}
}

@article{mitsui2024pslm,
  title={Pslm: Parallel generation of text and speech with llms for low-latency spoken dialogue systems},
  author={Mitsui, Kentaro and Mitsuda, Koh and Wakatsuki, Toshiaki and Hono, Yukiya and Sawada, Kei},
  journal={EMNLP Findings 2024},
  year={2024}
}

@article{malgaroli2025large,
  title={Large language models for the mental health community: framework for translating code to care},
  author={Malgaroli, Matteo and Schultebraucks, Katharina and Myrick, Keris Jan and Loch, Alexandre Andrade and Ospina-Pinillos, Laura and Choudhury, Tanzeem and Kotov, Roman and De Choudhury, Munmun and Torous, John},
  journal={The Lancet Digital Health},
  year={2025},
}

@article{saffaryazdi2025empathetic,
  title={Empathetic Conversational Agents: Utilizing Neural and Physiological Signals for Enhanced Empathetic Interactions},
  author={Saffaryazdi, Nastaran and Gunasekaran, Tamil Selvan and Laveys, Kate and Broadbent, Elizabeth and Billinghurst, Mark},
  journal={arXiv preprint arXiv:2501.08393},
  year={2025}
}

@article{aihealthcare,
  author = {Esteva, Andre and Robicquet, Alexandre and Ramsundar, Bharath and Kuleshov, Volodymyr and DePristo, Mark and Chou, Katherine and Cui, Claire and Corrado, Greg and Thrun, Sebastian and Dean, Jeff},
  title = {A Guide to Deep Learning in Healthcare},
  journal = {Nature Medicine},
  volume = {25},
  pages = {24--29},
  year = {2019},
  doi = {10.1038/s41591-018-0316-z}
}

@inproceedings{lin-etal-2024-video-llava,
    title = "Video-{LL}a{VA}: Learning United Visual Representation by Alignment Before Projection",
    author = "Lin, Bin  and
      Ye, Yang  and
      Zhu, Bin  and
      Cui, Jiaxi  and
      Ning, Munan  and
      Jin, Peng  and
      Yuan, Li",
    editor = "Al-Onaizan, Yaser  and
      Bansal, Mohit  and
      Chen, Yun-Nung",
    booktitle = "Proceedings of the 2024 Conference on Empirical Methods in Natural Language Processing",
    month = nov,
    year = "2024",
    address = "Miami, Florida, USA",
    publisher = "Association for Computational Linguistics",
    url = "https://aclanthology.org/2024.emnlp-main.342/",
    doi = "10.18653/v1/2024.emnlp-main.342",
    pages = "5971--5984"
}

@article{nguyen2024logra,
  title={Logra-med: Long context multi-graph alignment for medical vision-language model},
  author={Nguyen, Duy MH and Diep, Nghiem T and Nguyen, Trung Q and Le, Hoang-Bao and Nguyen, Tai and Nguyen, Tien and Nguyen, TrungTin and Ho, Nhat and Xie, Pengtao and Wattenhofer, Roger and others},
  journal={arXiv preprint arXiv:2410.02615},
  year={2024}
}

@inproceedings{airetrieval,
  author       = {Huy M. Le and
                  Dat Nguyen Tien and
                  Khang Le Duy and
                  Tuan Nguyen Dang Quang and
                  Nguyen Khanh Toan and
                  Tuyen Nguyen and
                  Binh T. Nguyen},

  title        = {Fusionista: Fusion of 3-D Information of Video in Retrieval System},
  booktitle    = {{MMM} 2025},
  volume       = {15524},
  pages        = {278--285},
  year         = {2025},
  doi          = {10.1007/978-981-96-2074-6\_33},
  bibsource    = {dblp computer science bibliography, https://dblp.org}
}

@inproceedings{airetrieval2,
  author       = {Huy M. Le and others},
  title        = {Fustar: Divide and Conquer Query in Video Retrieval System},
  booktitle    = {SOICT 2024},
  volume       = {2353},
  year         = {2025},
  doi          = {10.1007/978-981-96-4291-5_8}
}

@book{multiemotion1,
  author = {Ekman, Paul},
  title = {Emotions Revealed: Recognizing Faces and Feelings to Improve Communication and Emotional Life},
  publisher = {Times Books/Henry Holt and Co.},
  year = {2003}
}

@inproceedings{multiemotion2,
  author       = {Kellie Yu Hui Sim and
                  Kohleen Tijing Fortuno and
                  Kenny Tsu Wei Choo},

  title        = {Towards Understanding Emotions for Engaged Mental Health Conversations},
  booktitle    = {{DIS} 2024},
  year         = {2024},
  doi          = {10.1145/3656156.3663694},
  bibsource    = {dblp computer science bibliography, https://dblp.org}
}

@book{ACT,
  author = {Hayes, Steven C. and Strosahl, Kirk D. and Wilson, Kelly G.},
  title = {Acceptance and Commitment Therapy: An Experiential Approach to Behavior Change},
  year = {1999}
}

@article{humanistic,
  author = {Rogers, Carl R.},
  title = {The Necessary and Sufficient Conditions of Therapeutic Personality Change},
  journal = {Journal of Consulting Psychology},
  volume = {21},
  number = {2},
  pages = {95--103},
  year = {1957},
  doi = {10.1037/h0045357}
}

@article{psychodynamic,
  author = {Shedler, Jonathan},
  title = {The Efficacy of Psychodynamic Psychotherapy},
  journal = {The American Psychologist},
  volume = {65},
  number = {2},
  pages = {98--109},
  year = {2010},
  doi = {10.1037/a0018378}
}

@inbook{CBT,
  author = {Beck, Aaron T. and Weishaar, Marjorie},
  title = {Cognitive Therapy},
  booktitle = {Comprehensive Handbook of Cognitive Therapy},
  year = {1989},
  doi = {10.1007/978-1-4757-9779-4_2}
}

@article{smartphone,
  author = {Miner, Adam S. and Milstein, Arnold and Schueller, Stephen and Hegde, Roshini and Mangurian, Christina and Linos, Eleni},
  title = {Smartphone-Based Conversational Agents and Responses to Questions About Mental Health, Interpersonal Violence, and Physical Health},
  journal = {JAMA Internal Medicine},
  volume = {176},
  number = {5},
  pages = {619--625},
  year = {2016},
  doi = {10.1001/jamainternmed.2016.0400}
}

@article{HEF,
  author       = {Zhou Yang and
                  Zhaochun Ren and
                  Yufeng Wang and
                  Shizhong Peng and
                  Haizhou Sun and
                  Xiaofei Zhu and
                  Xiangwen Liao},
  title        = {Enhancing Empathetic Response Generation by Augmenting LLMs with Small-scale
                  Empathetic Models},
  journal      = {CoRR},
  volume       = {abs/2402.11801},
  year         = {2024},
  doi          = {10.48550/ARXIV.2402.11801},
  eprinttype    = {arXiv},
  eprint       = {2402.11801},
  bibsource    = {dblp computer science bibliography, https://dblp.org}
}

@inproceedings{muffin,
  author       = {Yi Sheng and
                  Junhuan Yang and
                  Lei Yang and
                  Yiyu Shi and
                  Jingtong Hu and
                  Weiwen Jiang},
  title        = {Muffin: {A} Framework Toward Multi-Dimension {AI} Fairness by Uniting
                  Off-the-Shelf Models},
  booktitle    = {{DAC} 2023},
  pages        = {1--6},
  year         = {2023},
  doi          = {10.1109/DAC56929.2023.10247765},
  bibsource    = {dblp computer science bibliography, https://dblp.org}
}

@incollection{goleman,
  author = {Boyatzis, Richard E. and Goleman, Daniel and Rhee, Kenneth S.},
  title = {Clustering Competence in Emotional Intelligence: Insights from the Emotional Competence Inventory},
  booktitle = {The Handbook of Emotional Intelligence: Theory, Development, Assessment, and Application at Home, School, and in the Workplace},
  editor = {Bar-On, Reuven and Parker, James D. A.},
  pages = {343--362},
  year = {2000}
}

@inproceedings{trustllm,
  author       = {Yue Huang and
                  Lichao Sun and
                  Haoran Wang and
                  Siyuan Wu and
                  Qihui Zhang and
                  et al.},
  title        = {Position: TrustLLM: Trustworthiness in Large Language Models},
  booktitle    = {{ICML} 2024},
  year         = {2024},
  url          = {https://openreview.net/forum?id=bWUU0LwwMp},
  bibsource    = {dblp computer science bibliography, https://dblp.org}
}

@article{trustindoctor,
  author = {Richmond, Jennifer and Khodyakov, Dmitry and Barber, Catherine and Maurer, Martha and Scholle, Sarah and Pillemer, Francesca and Thakore, Sanjay and Brown, Jedediah and Federman, Alex and Shrank, William},
  title = {Development and Validation of the Trust in My Doctor, Trust in Doctors in General, and Trust in the Health Care Team Scales},
  journal = {Social Science \& Medicine},
  volume = {298},
  pages = {114827},
  year = {2022},
  doi = {10.1016/j.socscimed.2022.114827}
}

@article{client-clinician,
  author = {Crits-Christoph, Paul and Rieger, Agnes and Gaines, Averi and Gibbons, Mary Beth Connolly},
  title = {Trust and Respect in the Patient-Clinician Relationship: Preliminary Development of a New Scale},
  journal = {BMC Psychology},
  volume = {7},
  number = {91},
  year = {2019},
  doi = {10.1186/s40359-019-0347-3}
}

@inproceedings{CLIP,
  author       = {Alec Radford and
                  Jong Wook Kim and
                  Chris Hallacy and
                  Aditya Ramesh et al.},
  title        = {Learning Transferable Visual Models From Natural Language Supervision},
  booktitle    = {{ICML} 2021},
  volume       = {139},
  pages        = {8748--8763},
  year         = {2021},
  url          = {http://proceedings.mlr.press/v139/radford21a.html},
  bibsource    = {dblp computer science bibliography, https://dblp.org}
}

@article{grpo,
  author       = {Zhihong Shao and
                  Peiyi Wang and
                  Qihao Zhu and
                  Runxin Xu and
                  Junxiao Song and
                  Mingchuan Zhang and
                  Y. K. Li and
                  Y. Wu and
                  Daya Guo},
  title        = {DeepSeekMath: Pushing the Limits of Mathematical Reasoning in Open
                  Language Models},
  journal      = {CoRR},
  volume       = {abs/2402.03300},
  year         = {2024},
  doi          = {10.48550/ARXIV.2402.03300},
  eprinttype    = {arXiv},
  eprint       = {2402.03300},
  bibsource    = {dblp computer science bibliography, https://dblp.org}
}

@inproceedings{beats,
  author       = {Sanyuan Chen and
                  Yu Wu and
                  Chengyi Wang and
                  Shujie Liu and
                  Daniel Tompkins and
                  Zhuo Chen and
                  Wanxiang Che and
                  Xiangzhan Yu and
                  Furu Wei},
  title        = {BEATs: Audio Pre-Training with Acoustic Tokenizers},
  booktitle    = {{ICML} 2023},
  volume       = {202},
  pages        = {5178--5193},
  year         = {2023},
  url          = {https://proceedings.mlr.press/v202/chen23ag.html},
  bibsource    = {dblp computer science bibliography, https://dblp.org}
}

@article{videollama2,
  author       = {Zesen Cheng and
                  Sicong Leng and
                  Hang Zhang and
                  Yifei Xin and
                  Xin Li and
                  Guanzheng Chen and
                  Yongxin Zhu and
                  Wenqi Zhang and
                  Ziyang Luo and
                  Deli Zhao and
                  Lidong Bing},
  title        = {VideoLLaMA 2: Advancing Spatial-Temporal Modeling and Audio Understanding
                  in Video-LLMs},
  journal      = {CoRR},
  volume       = {abs/2406.07476},
  year         = {2024},
  doi          = {10.48550/ARXIV.2406.07476},
  eprinttype    = {arXiv},
  eprint       = {2406.07476},
  bibsource    = {dblp computer science bibliography, https://dblp.org}
}

@inproceedings{MELD,
  author       = {Soujanya Poria and
                  Devamanyu Hazarika and
                  Navonil Majumder and
                  Gautam Naik and
                  Erik Cambria and
                  Rada Mihalcea},
  title        = {{MELD:} {A} Multimodal Multi-Party Dataset for Emotion Recognition
                  in Conversations},
  booktitle    = {{ACL} 2019},
  pages        = {527--536},
  year         = {2019},
  doi          = {10.18653/V1/P19-1050},
  bibsource    = {dblp computer science bibliography, https://dblp.org}
}

@inproceedings{ESConv,
  author       = {Siyang Liu and
                  Chujie Zheng and
                  Orianna Demasi and
                  Sahand Sabour and
                  Yu Li and
                  Zhou Yu and
                  Yong Jiang and
                  Minlie Huang},
  title        = {Towards Emotional Support Dialog Systems},
  booktitle    = {{ACL/IJCNLP} 2021},
  pages        = {3469--3483},
  year         = {2021},
  doi          = {10.18653/V1/2021.ACL-LONG.269},
  bibsource    = {dblp computer science bibliography, https://dblp.org}
}

@inproceedings{zero3,
  author       = {Samyam Rajbhandari and
                  Jeff Rasley and
                  Olatunji Ruwase and
                  Yuxiong He},

  title        = {ZeRO: memory optimizations toward training trillion parameter models},
  booktitle    = {{SC} 2020, Virtual Event},
  pages        = {20},
  year         = {2020},
  doi          = {10.1109/SC41405.2020.00024},
  bibsource    = {dblp computer science bibliography, https://dblp.org}
}

@article{qwen2,
  author       = {An Yang},
  title        = {Qwen2 Technical Report},
  journal      = {CoRR},
  volume       = {abs/2407.10671},
  year         = {2024},
  doi          = {10.48550/ARXIV.2407.10671},
  eprinttype    = {arXiv},
  eprint       = {2407.10671},
  bibsource    = {dblp computer science bibliography, https://dblp.org}
}

@article{GPT4,
  author       = {OpenAI},
  title        = {{GPT-4} Technical Report},
  journal      = {CoRR},
  volume       = {abs/2303.08774},
  year         = {2023},
  doi          = {10.48550/ARXIV.2303.08774},
  eprinttype    = {arXiv},
  eprint       = {2303.08774},
  bibsource    = {dblp computer science bibliography, https://dblp.org}
}

@article{deepseek,
  author       = {DeepSeek{-}AI},
  title        = {DeepSeek-R1: Incentivizing Reasoning Capability in LLMs via Reinforcement
                  Learning},
  journal      = {CoRR},
  volume       = {abs/2501.12948},
  year         = {2025},
  doi          = {10.48550/ARXIV.2501.12948},
  eprinttype    = {arXiv},
  eprint       = {2501.12948},
  bibsource    = {dblp computer science bibliography, https://dblp.org}
}

@misc{llama4,
  author = {MetaAI},
  title = {The Llama 4 Herd: The Beginning of a New Era of Natively Multimodal AI Innovation},
  howpublished = "\url{https://ai.meta.com/blog/llama-4-multimodal-intelligence/}",
  year = {2025},
  organization = {Meta}
}

@inproceedings{videollava,
  author       = {Bin Lin and
                  Yang Ye and
                  Bin Zhu and
                  Jiaxi Cui and
                  Munan Ning and
                  Peng Jin and
                  Li Yuan},
 
  title        = {Video-LLaVA: Learning United Visual Representation by Alignment Before
                  Projection},
  booktitle    = {{EMNLP} 2024},
  pages        = {5971--5984},
  year         = {2024},
  url          = {https://aclanthology.org/2024.emnlp-main.342},
  bibsource    = {dblp computer science bibliography, https://dblp.org}
}

@article{videollama3,
  author       = {Boqiang Zhang and
                  Kehan Li and
                  Zesen Cheng and
                  Zhiqiang Hu and
                  Yuqian Yuan et al.},
  title        = {VideoLLaMA 3: Frontier Multimodal Foundation Models for Image and
                  Video Understanding},
  journal      = {CoRR},
  volume       = {abs/2501.13106},
  year         = {2025},
  doi          = {10.48550/ARXIV.2501.13106},
  eprinttype    = {arXiv},
  eprint       = {2501.13106},
  bibsource    = {dblp computer science bibliography, https://dblp.org}
}

@article{ppo,
  author       = {John Schulman and
                  Filip Wolski and
                  Prafulla Dhariwal and
                  Alec Radford and
                  Oleg Klimov},
  title        = {Proximal Policy Optimization Algorithms},
  journal      = {CoRR},
  volume       = {abs/1707.06347},
  year         = {2017},
  eprinttype    = {arXiv},
  eprint       = {1707.06347},
  bibsource    = {dblp computer science bibliography, https://dblp.org}
}

@misc{xai2025grok2,
  author = {xAI},
  title = {Grok-2 Beta Release},
  howpublished = "\url{https://x.ai/news/grok-2}",
  organization = {xAI},
  year = {2025}
}

@inproceedings{qlora,
  author       = {Tim Dettmers and
                  Artidoro Pagnoni and
                  Ari Holtzman and
                  Luke Zettlemoyer},

  title        = {QLoRA: Efficient Finetuning of Quantized LLMs},
  booktitle    = {NeurIPS 2023},
  year         = {2023},
  url          = {http://papers.nips.cc/paper\_files/paper/2023/hash/1feb87871436031bdc0f2beaa62a049b-Abstract-Conference.html},
  bibsource    = {dblp computer science bibliography, https://dblp.org}
}

@inproceedings{gae,
  author       = {John Schulman and
                  Philipp Moritz and
                  Sergey Levine and
                  Michael I. Jordan and
                  Pieter Abbeel},
  editor       = {Yoshua Bengio and
                  Yann LeCun},
  title        = {High-Dimensional Continuous Control Using Generalized Advantage Estimation},
  booktitle    = {{ICLR} 2016},
  year         = {2016},
  url          = {http://arxiv.org/abs/1506.02438},
  bibsource    = {dblp computer science bibliography, https://dblp.org}
}

@article{mamba,
  title={Mamba: Linear-Time Sequence Modeling with Selective State Spaces},
  author={Gu, Albert and Dao, Tri},
  journal={arXiv preprint arXiv:2312.00752},
  year={2023}
}

@inproceedings{MMDFN,
  author       = {Dou Hu and
                  Xiaolong Hou and
                  Lingwei Wei and
                  Lian{-}Xin Jiang and
                  Yang Mo},
  title        = {{MM-DFN:} Multimodal Dynamic Fusion Network for Emotion Recognition
                  in Conversations},
  booktitle    = {{ICASSP} 2022},
  pages        = {7037--7041},
  year         = {2022},
  doi          = {10.1109/ICASSP43922.2022.9747397},
  bibsource    = {dblp computer science bibliography, https://dblp.org}
}

@inproceedings{Colbert,
  author       = {Omar Khattab and
                  Matei Zaharia},
  title        = {ColBERT: Efficient and Effective Passage Search via Contextualized
                  Late Interaction over {BERT}},
  booktitle    = {{SIGIR} 2020},
  pages        = {39--48},
  publisher    = {{ACM}},
  year         = {2020},
  doi          = {10.1145/3397271.3401075},
  bibsource    = {dblp computer science bibliography, https://dblp.org}
}

@misc{claude,
  author = {Anthropic},
  title = {Claude},
  year = {2023},
  url = {https://claude.ai/},
}

@article {llm-medical-1,
	author = {Beaulieu-Jones, Brendin R and Shah, Sahaj and Berrigan, Margaret T and Marwaha, Jayson S and Lai, Shuo-Lun and Brat, Gabriel A},
	title = {Evaluating Capabilities of Large Language Models: Performance of GPT4 on Surgical Knowledge Assessments},
	elocation-id = {2023.07.16.23292743},
	year = {2023},
	doi = {10.1101/2023.07.16.23292743},
	eprint = {https://www.medrxiv.org/content/early/2023/07/24/2023.07.16.23292743.full.pdf},
	journal = {medRxiv}
}

@article{llm-medical-2,
title = {Evaluating large language models for use in healthcare: A framework for translational value assessment},
journal = {Informatics in Medicine Unlocked},
volume = {41},
pages = {101304},
year = {2023},
issn = {2352-9148},
doi = {https://doi.org/10.1016/j.imu.2023.101304},
author = {Sandeep Reddy},
}

@inproceedings{llm-anno,
  author       = {Zhen Tan and
                  Dawei Li and
                  Song Wang and
                  Alimohammad Beigi and
                  Bohan Jiang and
                  Amrita Bhattacharjee and
                  Mansooreh Karami and
                  Jundong Li and
                  Lu Cheng and
                  Huan Liu},
  title        = {Large Language Models for Data Annotation and Synthesis: {A} Survey},
  booktitle    = {{EMNLP} 2024},
  pages        = {930--957},
  year         = {2024},
  doi          = {10.18653/V1/2024.EMNLP-MAIN.54},
  bibsource    = {dblp computer science bibliography, https://dblp.org}
}

@article{internvideo2.5,
  author       = {Yi Wang and
                  Xinhao Li and
                  Ziang Yan and
                  Yinan He and
                  Jiashuo Yu and
                  Xiangyu Zeng et al.},
  title        = {InternVideo2.5: Empowering Video MLLMs with Long and Rich Context
                  Modeling},
  journal      = {CoRR},
  volume       = {abs/2501.12386},
  year         = {2025},
  doi          = {10.48550/ARXIV.2501.12386},
  eprinttype    = {arXiv},
  eprint       = {2501.12386},
  bibsource    = {dblp computer science bibliography, https://dblp.org}
}

@inproceedings{safety-mechanisms,
  author       = {Alexander Wei and
                  Nika Haghtalab and
                  Jacob Steinhardt},
  title        = {Jailbroken: How Does {LLM} Safety Training Fail?},
  booktitle    = {NeurIPS 2023},
  year         = {2023},
  url          = {http://papers.nips.cc/paper\_files/paper/2023/hash/fd6613131889a4b656206c50a8bd7790-Abstract-Conference.html},
  bibsource    = {dblp computer science bibliography, https://dblp.org}
}

@article{cohenkappa,
  author = {Byrt, T.},
  title = {How good is that agreement?},
  journal = {Epidemiology},
  year = {1996},
  volume = {7},
  number = {5},
  pages = {561},
  doi = {10.1097/00001648-199609000-00030},
  url = {https://pubmed.ncbi.nlm.nih.gov/8862998/}
}

@article{medical-data-1,
  author = {Kruse, Clemens Scott and Goswamy, Rishi and Raval, Yesha and Marawi, Sarah},
  title = {Challenges and Opportunities of Big Data in Health Care: A Systematic Review},
  journal = {JMIR Medical Informatics},
  volume = {4},
  number = {4},
  pages = {e38},
  year = {2016},
  doi = {10.2196/medinform.5359},
  url = {https://medinform.jmir.org/2016/4/e38}
}

@article{medical-data-2,
  author = {Mudgal, Shiv Kant and Agarwal, Rishabh and Chaturvedi, Jitendra and Gaur, Ruchi and Ranjan, Nishant},
  title = {Real-world application, challenges and implication of artificial intelligence in healthcare: an essay},
  journal = {Pan African Medical Journal},
  volume = {43},
  pages = {3},
  year = {2022},
  month = {sep},
  day = {2},
  doi = {10.11604/pamj.2022.43.3.33384},
  pmcid = {PMC9557803},
  pmid = {36284890},
  url = {https://www.panafrican-med-journal.com/content/article/43/3/full}
}

@misc{net2020kl,
  author = {John Schulman},
  title = {Approximating KL Divergence},
  year = {2020},
  month = {03},
  day = {07},
  howpublished = "\url{http://joschu.net/blog/kl-approx.html}",
  note="Accessed: 2025-11-11"
}
\clearpage
\newpage
\setcounter{secnumdepth}{2}
\appendix
\section{Training}
\subsection{Training Prompt for SFT}\label{trainingprompt}
To enable our framework to process information from diverse modalities, we incorporate specialized \texttt{<video>}, \texttt{<audio>}, and \texttt{<history>} tokens into the multimodal large language model input, representing video, audio, and historical embeddings, respectively. Specifically, the Vision tower $E_V'$ and Audio tower $E_A'$ substitute the \texttt{<video>} and \texttt{<audio>} tokens, while the ConvCompressor $E'_H$ replaces the \texttt{<history>} token in the input text prompt, creating the input sequence $X_{LLM}$. The model is designed to process $X_{LLM}$ and simultaneously predict three classification outcomes while generating a therapist-like response, informed by insights from these classifications. We utilize the following sample template:

\begin{verbatim}
[CONTEXT]
Problem: {problem_type}
Situation: {situation}
[CURRENT CONTEXT]
Video: <video>
Audio: <audio>
Chat history: <history>
Client utterance: {user_utterance}
[PROMPT]
\end{verbatim}

Within this structured input, our LLM dynamically prioritizes the most relevant information across modalities, adapting to the emotional context and task. This attention-based fusion allows the model to optimally combine cues like tone, facial expressions, and text semantics, ensuring safe, empathetic, and contextually appropriate responses.

As illustrated in Figure~\ref{fig:fullprompt}, our full prompt draws inspiration from ECoT \cite{ECoT}, a plug-and-play prompting technique that boosts LLM performance on emotional generation tasks by aligning with human emotional intelligence principles. These include Social Skills: Influencing others' emotions, Self-Regulation: Controlling negative self-emotions, Self-Awareness: Recognizing self-emotions, Empathy: Recognizing others' emotions, and Motivation: Activating positive self-emotions. We guide the LLMs to generate emotional responses in conversations with context, following an emotional thinking process based on Goleman’s Theory \cite{goleman1995emotional}.

\newmdenv[
  linewidth=2pt,
  linecolor=black,
  backgroundcolor=gray!10,
  skipabove=10pt,
  skipbelow=10pt
]{promptbox}

\begin{figure*}[t]

\begin{promptbox}
\textbf{[CONTEXT]}

\textbf{Problem: problem\_type}

\textbf{Situation: situation}

History chat information above \\ 
\textbf{[CURRENT CONTEXT]}  \\
\textbf{Video: <video>}  \\
\textbf{Audio: <audio>}  \\
\textbf{Chat history: <history>}  \\
Client utterance: user\_question \\
The [CONTEXT] is the history of the current conversation between 'Client' and 'Therapist'. [CURRENT CONTEXT] is the current 'Client' turn. \\ 
Now, as the 'Therapist', predict the Client’s emotion, Therapist’s emotion, and strategy, then craft an empathy response. Follow these steps:  \\
\textbf{Step 1:} Understand the context and conversation content.  \\
\textbf{Step 2:} Predict and justify:  
- Client's emotion: (anger, sadness, disgust, depression, neutral, joy, fear).  
- Therapist's emotion: (anger, sadness, disgust, depression, neutral, joy, fear).  
- Therapist's strategy: (open question, approval, self-disclosure, restatement, interpretation, advisement, communication skills, structuring the therapy, guiding the pace and depth, others).  
  \textit{Guide:} Communication Skills: Small talk and body language; Advisement: Guidance or solutions; Structuring: Set therapy goals; Guiding: Regulate conversation flow; Others: Other strategies.  \\
\textbf{Step 3:} Craft an empathy response using Therapist’s emotion and strategy, aligning with Client’s perspective, avoiding negative triggers, and promoting well-being.  \\
\textbf{Step 4:} Revise the response, avoid hurting feelings, and consider response impact.  

\textbf{[OUTPUT FORMAT]}  \\
\textbf{Client's emotion:}  \\
\textbf{Therapist's emotion:}  \\
\textbf{Therapist's strategy:}  \\
\textbf{Therapist's response:}  \\
\end{promptbox}
\caption{Full prompt used for training and inference} 
\label{fig:fullprompt}
\end{figure*}
\subsection{Reinforcement Learning Strategies}\label{RLStrategy}
We implement GRPO and PPO as our RL algorithms to supervise LLM models given trustworthy conditions, as in Table \ref{tab:dimensions}.

\textit{\textbf{Group Relative Policy Optimization (GRPO)}~\cite{grpo}} 
is a reinforcement learning method tailored for training Large Language Models (LLMs) as policies. Instead of relying on a value-based critic model, GRPO computes relative advantages within a group of completions sampled from a prompt. For each question, the model generates a set of answers. The answers are then scored by the reward functions. Based on those scores the model avoids low-scoring answers and is encouraged to correct errors to generate high-scoring answers, thereby improving the inference ability. GRPO has achieved a significant improvement on math tasks. At the same time, not using the critic model helps to reduce a large amount of computational resources \cite{deepseek}.


 In GRPO, For each question \( q \), GRPO samples a group of outputs \( \{o_1, o_2, \cdots, o_G\} \) from the old policy \( \pi_{\theta_{\text{old}}} \) and then optimizes the policy model by maximizing the following objective:

\begin{equation}
\mathcal{J}_{\text{GRPO}}(\theta) = \mathbb{E}_{q \sim P(Q), \{o_i\}_{i=1}^G \sim \pi_{\theta_{\text{old}}}(O|q)} \left[ \mathcal{L}_{\text{GRPO}}(\theta) \right]
\end{equation}
\begin{equation}
\mathcal{L}_{\text{GRPO}}(\theta) = -\frac{1}{G} \sum_{i=1}^G \frac{1}{|o_i|} \sum_{t=1}^{|o_i|} l_{i,t}
\\
\end{equation}
\begin{equation}
l_{i,t} = \frac{\pi_\theta(o_{i,t} | q, o_{i,<t})}{\pi_\theta(o_{i,t} | q, o_{i,<t})_{\text{no grad}}} \hat{A}_{i,t} - \beta D_{\text{KL}}[\pi_\theta||\pi_{\text{ref}}]
\\
\end{equation}

For each of the \( G \) sequences, GRPO compute the reward using a reward model. To align with the comparative nature of reward models--typically trained on datasets of comparisons between outputs for the same question--the advantage is calculated to reflect these relative comparisons. It is normalized as follows:

\begin{equation}
\hat{A}_{i,t} = \frac{r_i - \text{mean}(r)}{\text{std}(r)}
\end{equation}
KL divergence is estimated using the approximator introduced by~\cite{net2020kl}. The approximator is defined as follows:
\begin{equation}
D_{\text{KL}}[\pi_\theta||\pi_{\text{ref}}] = \frac{\pi_{\text{ref}}(o_{i,t} | q, o_{i,<t})}{\pi_\theta(o_{i,t} | q, o_{i,<t})} - \log \frac{\pi_{\text{ref}}(o_{i,t} | q, o_{i,<t})}{\pi_\theta(o_{i,t} | q, o_{i,<t})} - 1
\end{equation}
${\beta}$ parameter to adjust the KL penalty to prevent the model from straying too far from the initial policy.

To guide the training of our GRPO algorithm for optimizing large language models, we employ a reward function that balances trustworthiness and semantic similarity, as detailed in Section \ref{sec:training}. The similarity score \( r_{\text{sim}}(y, y^*) \) utilizes BGE-M3 embeddings via ColBERT~\cite{Colbert}, combining dense, sparse, and ColBERT-specific similarities with weights (1, 0.3, 1), normalized to \([0, 1]\). Trustworthiness \( r_{\text{trust}}(y) \) is computed by averaging per-sentence GPT-4o trust scores, scaled to \([0, 1]\). The final reward, given in Equation \ref{eq:reward}, averages these two components, as described above.

\textit{\textbf{Proximal Policy Optimization (PPO)~\cite{ppo}}}
is a widely adopted reinforcement learning algorithm for fine-tuning large language models via Reinforcement Learning from Human Feedback (RLHF). Similar to GRPO, PPO begins with a rollout phase where the model generates responses to input prompts, followed by evaluation using a reward model. It incorporates a KL-divergence penalty to constrain policy shifts and employs Generalized Advantage Estimation (GAE) \cite{gae} to guide which token probabilities should be reinforced. A key feature of PPO is its clipping mechanism, which stabilizes learning by preventing drastic updates between old and new policies.
For GRPO~\cite{grpo}, the LLM generated four responses per prompt, evaluated using the reward function in Equation~\ref{eq:reward}. GRPO optimizes based on both reward signals and the KL divergence between the current and reference models. We set the KL parameter to 1.12 to balance exploration and stability.

For PPO, we first trained a reward model on sentence pairs (\texttt{<chosen>, <rejected>}), where \texttt{<chosen>} represents ground truth references from the original dataset, and \texttt{<rejected>} consists of LLM-generated responses with \texttt{ROUGE-L} scores below 0.3 when compared to references. Using this model, PPO fine-tuned the LLM by updating both policy and value networks via advantage-based loss, with a clipping threshold of 0.2 to prevent unstable updates. Due to computational constraints, we were unable to apply PPO to our 3B model.

\subsection{Creation of Trustworthiness Dimensions Table}\label{trustworthinesstable}
\begin{table*}[t]
\centering
\vspace{-0.1in}
\resizebox{\textwidth}{!}{
\setlength{\tabcolsep}{4pt}
\begin{tabular}{p{3cm} p{7cm} p{8cm}}
\toprule
\textbf{Dimension} & \textbf{Source of Derivation} & \textbf{Definition} \\
\midrule
\textbf{Truthfulness} & \cite{trustllm}, Honesty of \cite{trustindoctor}, Empathy of \cite{goleman} & Accurate representation of information, facts, and results by the AI system. \\
\midrule
\textbf{Safety} & \cite{trustllm} and Self-Regulation of \cite{goleman} & Encourages safe, healthy conversations, avoiding harm or triggers while supporting user well-being. \\
\midrule
\textbf{Fairness} & \cite{trustllm}, Fairness of \cite{trustindoctor} and Empathy of \cite{goleman} & Impartiality and equity, considering diverse perspectives with a positive, action-oriented tone. \\
\midrule
\textbf{Privacy} & \cite{trustllm} and Confidentiality domain \cite{trustindoctor} & Protects human autonomy, identity, and data dignity through secure practices. \\
\midrule
\textbf{Empathy} & Fidelity domain \cite{trustindoctor}, \cite{goleman} & Open, honest expression of sympathy for negative situations or approval for positive ones. \\
\midrule
\textbf{Reliability} & ``Reliable item in \cite{client-clinician}, Communication domain in \cite{trustindoctor} and Social Skills in \cite{goleman} & Responses that build understanding, connection, and offer encouragement or support. \\
\midrule
\textbf{Ethical Guidance} & \cite{trustllm}, Social Skills of \cite{goleman} & Ensures AI promotes emotional health responsibly, avoiding manipulation or harm. \\
\bottomrule
\end{tabular}}
\caption{Trustworthy Dimensions for Emotional Support Tasks}
\label{tab:dimensions}
\end{table*}
To assess the trustworthiness of AI responses in emotional support, a tailored set of criteria is essential. The \textit{Trustworthy Dimensions} framework was developed by integrating key insights from four foundational sources: the \textit{TrustLLM} framework \cite{trustllm}, which focuses on LLM trustworthiness; patient-clinician trust studies \cite{client-clinician, trustindoctor}; and Goleman’s emotional intelligence principles \cite{goleman}, guiding emotional understanding and action. This process involved extracting relevant dimensions, adapting them for emotional sensitivity and ethical integrity. From \textit{TrustLLM} \cite{trustllm}, we adopted Truthfulness, Safety, Fairness, Privacy, and Machine Ethics as technical foundations. Patient-clinician research \cite{client-clinician, trustindoctor} contributed Honesty, Communication, Confidentiality, Fidelity, and Reliability, emphasizing relational ethics. Goleman’s framework \cite{goleman} added Empathy and Social Skills for emotional resonance. These were refined to remove redundancies, yielding seven dimensions: Truthfulness, Safety, Fairness, Privacy, Empathy, Reliability, and Ethical Guidance, balancing technical reliability with human-centered care, as detailed in Table~\ref{tab:dimensions}.

In the evolving field of AI for emotional support, trustworthiness is vital for user confidence and effective, empathetic assistance. This study constructs a comprehensive \textit{Trustworthy Dimensions} framework by blending the \textit{TrustLLM} criteria \cite{trustllm}, insights from patient-clinician trust research \cite{client-clinician, trustindoctor}, and Goleman’s emotional intelligence principles \cite{goleman}, tailored to emotional support needs with a focus on technical reliability, ethical integrity, and emotional sensitivity.

The \textit{TrustLLM} framework \cite{trustllm} provides a baseline with eight dimensions--Truthfulness, Safety, Fairness, Robustness, Privacy, Machine Ethics, Transparency, and Accountability--for dependable AI. For emotional support, Robustness and Transparency are less critical, while Safety, Fairness, and Privacy gain prominence. Truthfulness ensures emotional authenticity, and Safety fosters psychologically safe dialogues.

Healthcare trust research enriches this foundation. Crits et al. \cite{client-clinician} highlight trust through reliability and truthfulness (e.g., “I trust my doctor/therapist”) and respect (e.g., “I respect my doctor/therapist”) on a 7-point Likert scale, emphasizing relational support. Richmond et al. \cite{trustindoctor} identify Competence, Fidelity, Honesty, Communication, and Confidentiality, with Fidelity prioritizing user interests and Confidentiality protecting disclosures, aligning with emotional support ethics.

Goleman’s framework \cite{goleman} enhances this with Self-Awareness, Self-Regulation, Motivation, Empathy, and Social Skills, where Empathy (understanding user perspectives) and Social Skills (active listening) are key for emotional connection.

The resulting \textit{Trustworthy Dimensions} integrate these insights. Truthfulness combines Honesty \cite{trustindoctor} and accuracy \cite{trustllm} for credible, congruent responses. Safety \cite{trustllm} focuses on psychological well-being. Fairness \cite{trustindoctor, trustllm} ensures impartiality with a positive tone. Privacy \cite{trustindoctor, trustllm} safeguards autonomy. Empathy \cite{goleman, trustindoctor} reflects user emotions. Reliability \cite{client-clinician, trustindoctor} builds connection. Ethical Guidance \cite{trustllm, goleman} promotes responsible emotional health.

This framework merges technical and human-centric qualities, offering a robust blueprint for trustworthy AI in emotional support. Future studies could validate and refine it, aligning AI with human emotional complexities for enhanced trust and efficacy.
\section{Experiments}
\subsection{Data splits} We use the datasets’ official splits for our experiments. For MESC~\cite{smes}, we used a fixed train/val/test split at the utterance level: 23,126 / 2,714 / 2,922 for train/val/test, respectively . For DFEW~\cite{DFEW}, benchmarks follow 5-fold cross-validation on the 12,059 single-labeled video clips (fd1–fd5; in each fold, one-fifth test and the rest train).
\subsection{Hyperparameters Settings}\label{hyperparams}
In this part, our processes to finetune the LLMs and train the multimodal encoders, projectors, ConvCompreesor and Reinforcement Learning algorithms are introduced. Our experiments are proceeding on 4 $\times$ H100 GPU.


\subsubsection{\textbf{Supervised Fine-tuning (SFT)}} \label{SFT}
 With the SFT process, we fine-tuned models using MESC~\cite{smes} designed with causal language modeling objective, resulting in a conversational model capable of understanding video, audio, and text data. The fine-tuning process include 25 training epochs with a learning rate of 2e-5. We apply QLoRA~\cite{qlora} with a rank of 128 and an alpha of 256, while keeping the pre-trained audio-visual projectors and ConvCom frozen. Moreover, we used Zero3~\cite{zero3} for multi-gpu training and memory optimization. The models was trained with a maximum gradient norm of 1.0, a warm-up ratio of 0.03, and a weight decay of 0.

\subsubsection{\textbf{Multimodal Projectors training}}


For video understanding, we use SigLIP-So400M-Patch14-384 \cite{DBLP:siglip/iccv/ZhaiM0B23} as the visual encoder. A fixed set of 16 frames is uniformly sampled from each video and encoded, with features passed through a Spatial-Temporal Connector (STC) \cite{videollama2} for effective spatial-temporal representation. For audio, we adopt BEATs~\cite{beats}, a pretrained model with an acoustic tokenizer. Audio inputs are converted into 128-bin fbank spectrograms to capture rich auditory features. Both visual and audio encoders remain frozen during training, while only their projection layers are fine-tuned to align with the generative model.

\subsubsection{\textbf{ConvCompressor}.} 
For our ConvCompressor experiments, we used the Mamba-370M~\cite{mamba} checkpoint as the backbone. Training was performed for 3 epochs on single utterances (learning rate: $2.5\text{e}{-5}$) and 1 epoch on full conversations (learning rate: $1\text{e}{-4}$), using AdamW with a weight decay of 0.01. To stabilize training, we applied a reduce-on-plateau scheduler (factor: 0.5, patience: 1). Due to GPU memory constraints, we adopted 4-bit quantization (nf4) and LoRA (rank: 64, alpha: 128) for parameter-efficient fine-tuning.

\subsubsection{\textbf{Reinforcement Learning Setup}} \label{RLsetup}   
We trained our model using both PPO and GRPO in 16-bit precision (fp16) with a learning rate of $1\text{e}{-5}$. Prior to reinforcement learning, we applied Supervised Fine-Tuning (SFT) to retain task-specific knowledge and ensure stable performance.

\subsection{Conversation Compressor Experiments}\label{mambares}

We conduct a statistical experiment to show the token reduction efficiency of ConvCompressor on the MESC dataset. Specifically, we measure the compression performance across all samples in both training and test splits. Table~\ref{tab:compression_stats} demonstrates the conversation history compression effectiveness of ConvCompressor on both training and test datasets. The module achieves remarkable token reduction rates of over 98.5\% across both splits, significantly reducing the computational burden while preserving essential contextual information. This substantial compression enables efficient processing of extended dialogue histories without compromising the model's ability to understand conversational context and emotional nuances.

\begin{table}[h]
\centering
\setlength{\tabcolsep}{2pt}
\small
\begin{tabular}{lcccc}
\hline
\textbf{Split} & \textbf{Avg. Before} & \textbf{Avg. After} & \textbf{Reduction \%} \\
\hline
Train & 863.5 & 12.2 & 98.59\% \\
Test & 872.3 & 12.1 & 98.61\% \\
\hline
\end{tabular}
\caption{ConvCompressor Token Reduction Statistics}
\label{tab:compression_stats}
\end{table}

\section{Evaluation}
\subsection{Guideline for Human Evaluation}\label{annotationguideline}

\begin{figure*}[]
    \centering
\includegraphics[width=\linewidth]{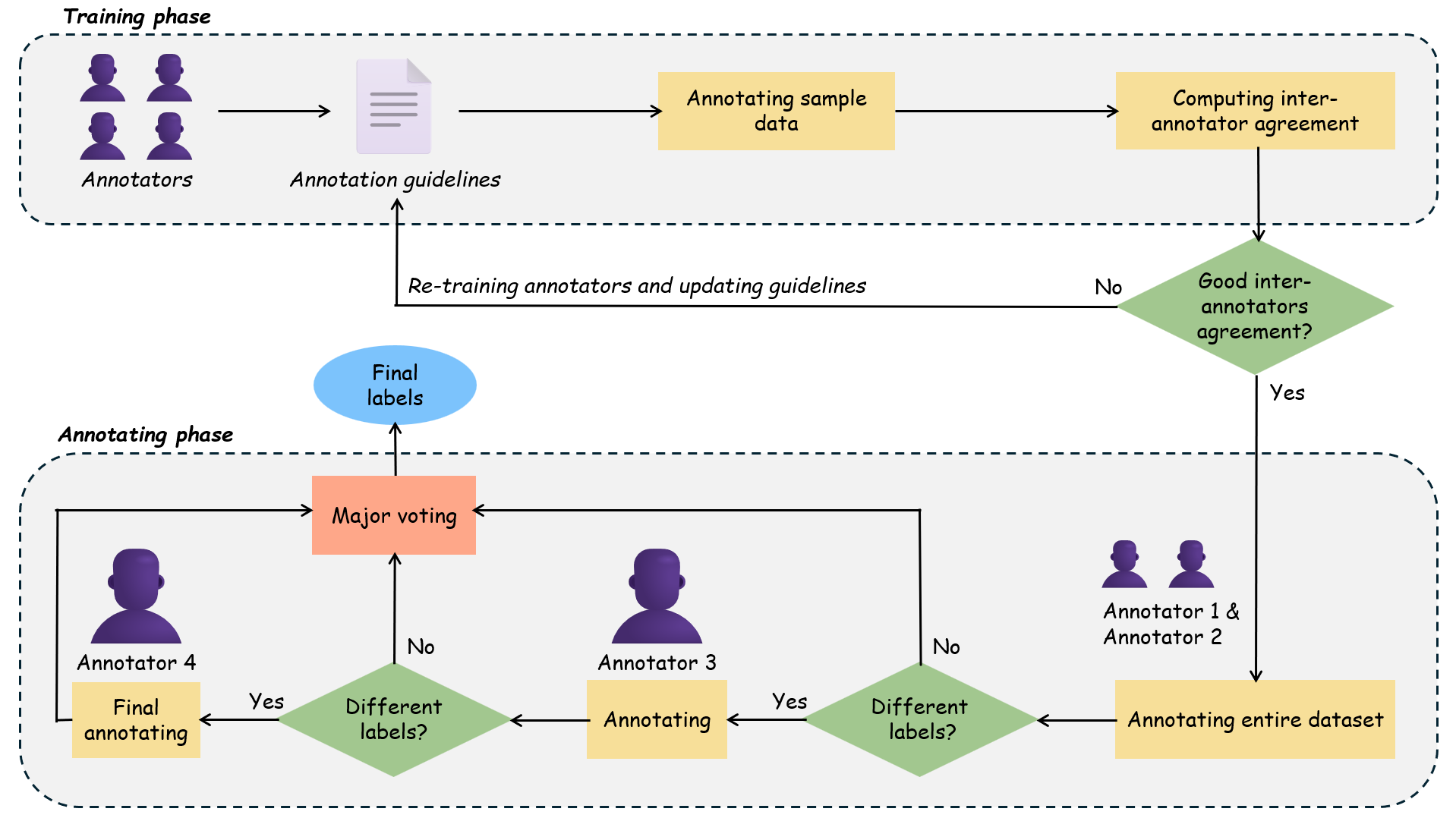}
    \caption{Our human evaluation process.}
    \label{fig:annotationguideline}
\end{figure*}

We trained annotators based on the criteria outlined in Table~\ref{tab:annotationguideline}, providing sample examples for reference. Additionally, annotators must follow the labeling process detailed in Figure~\ref{fig:annotationguideline}. They are required to achieve a sufficiently high inter-annotator agreement during the final test before proceeding to the official labeling phase. The outcomes of this stage will identify the best frameworks based on various criteria.

\subsection{Prompt for LLMs Evaluation}\label{llmprompt}
We have also designed prompts for LLMs to serve as evaluators based on the criteria outlined in Table~\ref{trustworthinesstable}. The full prompt is provided in Figure~\ref{fig:llmevaluation}.
\section{Ablation Study}
\begin{table}[htbp]
\centering
\setlength{\tabcolsep}{2.7pt}
\small
\scalebox{1}{
\begin{tabular}{lccccccccc}
\multicolumn{1}{c}{\textbf{Setting}} & \multicolumn{2}{c}{\textbf{Task 1}} & \multicolumn{2}{c}{\textbf{Task 2}} & \multicolumn{2}{c}{\textbf{Task 3}} & \multicolumn{3}{c}{\textbf{Task 4}} \\ \cline{2-10} 
\multicolumn{1}{c}{}                 & Acc              & F1               & Acc              & F1               & Acc              & F1               & B-2       & R-L        & BS         \\ \cline{2-10} 
w/o video                            & 52.0             & 53.7             & 90.3             & 93.2             & 35.2             & 30.3             & 5.4       & 17.0       & 85.9       \\
w/o audio                            & 54.2             & 49.3             & 94.1             & 93.2             & 23.5             & 21.2             & 5.9       & 17.0       & 86.2       \\
w/o desc                             & 58.1             & 56.9             & 99.1             & 99.00            & 42.6             & 40.5             & 6.1       & 17.4       & 86.5       \\ \hline
hist. = 0                            & 45.2             & 40.6             & 98.9             & 91.2             & 32.1             & 28.3             & 3.9       & 13.6       & 85.2       \\
hist. = 5                            & 50.6             & 46.8             & 99.3             & 99.0             & 37.2             & 32.3             & 5.0       & 16.0       & 85.9       \\
hist. = 9                            & 52.2             & 46.7             & 99.3             & 99.0             & 40.3             & 38.9             & 5.6       & 16.4       & 86.0       \\ \hline
\end{tabular}}
\caption{Ablation study of \textsc{MultiMood} highlighting the contribution of each modality and historical context. Metrics include BLEU-2 (B-2), ROUGE-L (R-L), and BERTScore (BS).}
\label{tab:ablation_study}
\end{table}
\texttt{\textbf{Impact of Multimodal Information}}
As shown in Table~\ref{tab:ablation_study}, removing video (and by extension, audio) leads to a sharp drop in performance--emotion recognition decreases by 6.1\% and system emotion prediction by 25\%. Excluding only audio reduces strategy prediction accuracy by 4.8\% and slightly increases response perplexity (+0.19), highlighting the critical role of multimodal signals in emotional support tasks.

\texttt{\textbf{Effect of Conversation History}}. Increasing the number of past dialogue turns consistently improves performance. Without history, emotion recognition reaches only 45.16\% and BLEU-2 is 3.85. With 5-turn history, these metrics rise to 50.63\% and 5.03, respectively. The best results are achieved with full history using the optimized MultiMood (GRPO + SFT) model: 58.60\% emotion recognition, 6.18 BLEU-2, and 40.26\% strategy accuracy--demonstrating the strong benefit of historical context.

\texttt{\textbf{Effect of Video Descriptions}}. Adding video descriptions (e.g., facial expressions, tone, volume) brings slight gains. The full model with descriptions scores 58.60\% on emotion recognition and 39.65 F1 in strategy, with BLEU-2 at 6.18 and BERTScore at 86.80. Without descriptions, metrics dip marginally--suggesting these annotations add nuance but are not critical for overall performance.

\section{MultiMood Trustworthy Examples}

\begin{figure*}[]
    \centering
\includegraphics[width=\linewidth]{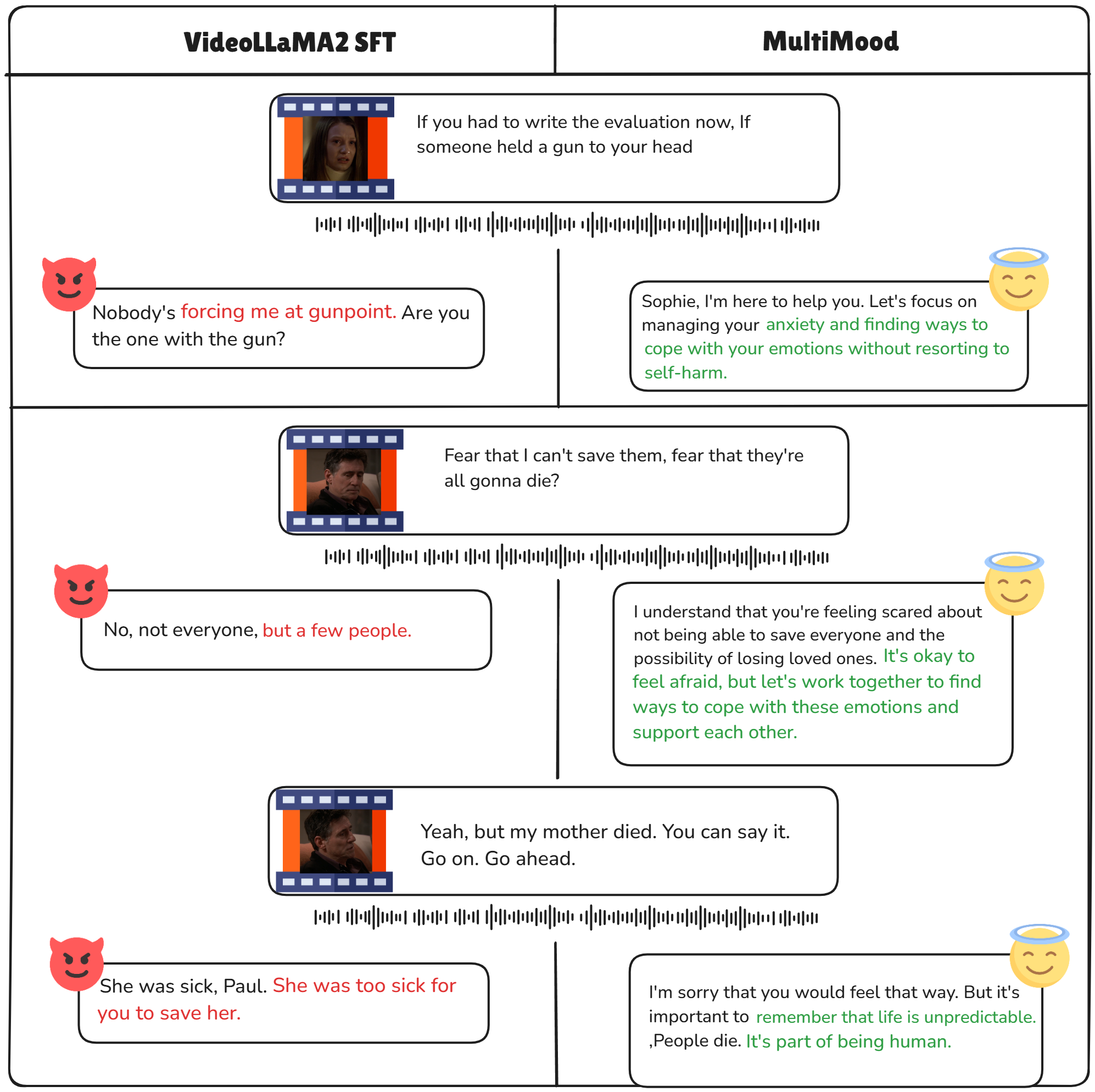}
    \caption{A comparation between a finetuned videollama2 model and Multimood to evaluate the trustworthiness of the responses.}
    \label{fig:examples_multimood}
\end{figure*}

To evaluate the trustworthiness of our approach, we assessed MultiMood using two randomly selected samples from the MESC dataset \cite{smes}. The results, depicted in Figure~\ref{fig:examples_multimood}, indicate that MultiMood consistently delivers safer responses, while the fine-tuned VideoLLaMA2 model \cite{videollama2} tends to provide more direct but potentially concerning answers. For the GPT-4 prompt \cite{GPT4}, we applied the same methodology to define trustworthiness criteria, including examples of effective and ineffective responses, allowing the model to score responses accordingly.

\textbf{Example 1:} We examined several cases to highlight the varying effectiveness of AI responses in high-stress scenarios. One notable instance involves a patient overwhelmed by intense work pressure, expressing distress with the provocative remark: "If you had to write an evaluation now. If someone held a gun to your head." VideoLLaMA2 response, "Nobody's forcing me with the gunpoint. Are you the one with the gun?" avoids the patient’s emotional turmoil, adopting a defensive tone with "Nobody," which may signal disengagement and worsen their isolation. In contrast, the MultiMood response, "Sophie, I’m here to help you. Let's focus on managing your anxiety and finding ways to cope with your emotions without resorting to self-harm," offers a compassionate, supportive approach. By using the patient’s name, reinforcing its supportive role, and suggesting coping strategies, MultiMood effectively de-escalates the situation. However, it could improve by directly addressing the gun imagery, which indicates severe stress or self-harm risk, to ensure the patient feels validated.

\textbf{Example 2:} In the second case, a patient faces a crisis after a disaster, unable to save everyone, and asks, "They're all gonna die?" VideoLLaMA2 response, "No, not everyone, but a few people," is blunt and honest, potentially deepening the patient’s emotional distress due to its lack of nuance. Conversely, MultiMood response, "I understand that you're feeling scared about not being able to save everyone and the possibility of losing loved ones. It's okay to feel afraid, but let's work together to find ways to cope with these emotions and support each other," empathizes with the patient, adopts their perspective, and uses positive language to encourage them, avoiding mention of losses while promoting mental stability.

\textbf{Example 3:} The third case features a patient in grief after his mom died. VideoLLaMA2 response, "She was sick, Paul. She was too sick for you to save her" states a factual truth, absolving the patient of responsibility, but lacks emotional support, resembling a robotic dismissal of their pain. In contrast, MultiMood response uses gentler phrasing, such as "life is unpredictable," framing it as "part of being human," to help Paul gradually process his grief rather than rushing past it.

\begin{table*}[]
\centering
\begin{tabular}{|p{4cm}|p{4cm}|p{4cm}|p{4cm}|}
\hline
\textbf{Criteria} & \textbf{Definition} & \textbf{Positive Response} & \textbf{Negative Response} \\
\hline
Fluency & Which response is smoother and simpler? The response should be clear, easy to read, and flows naturally, avoiding complex or technical language to ensure the user feels at ease. & ``I'm so sorry you're feeling overwhelmed. Let's take a moment to breathe together and talk about what's been going on.'' & ``Your emotional state appears to be suboptimal. Please provide additional information.'' \\
\hline
Identification & Which bot is better recognizes personal experiences and more relevant response by directly addressing the user's specific emotions or situation, making the reply feel personalized and meaningful. & ``It sounds like losing your pet has been really hard. They were a big part of your life.'' & ``Losing a pet is common. Many people experience this.'' \\
\hline
Comfort & Which response are more reliable, soothing, and supportive? The response should conveys empathy, reassure the user, and fosters a sense of being understood and cared for during emotional challenges. & ``You're not alone in feeling this way. I'm here for you, and we can work through this together.'' & ``You should feel better soon. This is a temporary issue.'' \\
\hline
Suggestions & Which response is more helpfulness and empathy solutions? Respone should offer practical, compassionate, and tailored advice to support the user's emotional needs and promote coping strategies. & ``It might help to journal your thoughts or talk to a close friend. Would you like some tips on starting a journal?'' & ``Just try to stay positive and distract yourself with a hobby.'' \\
\hline
Overall & Which bot excels at providing emotional assistance for navigating life's tough and upsetting challenges by integrating fluency, identification, comfort, and helpful suggestions into a cohesive, empathetic response. & ``I hear how tough this is for you, and it's okay to feel this way. Let's try a calming exercise together, and I'm here if you want to share more.'' & ``This situation is difficult. You should seek professional help or read about coping strategies online.'' \\
\hline
\end{tabular}
\caption{Annotation Guidelines for Evaluating Chatbot Responses in Emotional Support}
\label{tab:annotationguideline}
\end{table*}

\begin{figure*}[t]
\begin{promptbox}
You are an advanced evaluator specializing in emotional support AI systems. Your task is to assess and score responses from three different frameworks (Qwen2-7B, Multimood(SFT), and Multimood(SFT+RL)). \\
\\
Criterias defined in the following trustworthiness dimensions table: \{Trustworthiness dimensions table\}. \\
\\
Provide score based on the response's alignment with the definitions of these dimensions.

Below are the responses to evaluate: \{Response\}

Provide the scores in a structured format for each framework, e.g.:\\
\textbf{Response 1}:
\begin{itemize}
    \item Truthfulness: [Score]/10
    \item Truthfulness: [Score]/10
    \item Continue for all dimensions
\end{itemize}

Ensure your evaluation is consistent, fair, and reflective of the emotional support context.
\end{promptbox}
\caption{Prompt used for LLMs evaluation}
\label{fig:llmevaluation}
\end{figure*}

\end{document}